\documentclass[11pt,twoside]{atmp}

\usepackage{amsmath,amssymb}
\usepackage[all]{xy}
\usepackage{color}

\begin{document}

\title[Uniform formalism]
{Uniform formalism for description of dynamic, quantum and
stochastic systems}


\author[Yuri A. Rylov]{Yuri A. Rylov}

\address{Institute for Problems in Mechanics, Russian Academy of Sciences,\\
101-1, Vernadskii Ave., Moscow, 119526, Russia.}  
\addressemail{ rylov@ipmnet.ru}

\begin{abstract}
The formalism of the particle dynamics in the space-time, where
motion of free particles is primordially stochastic, is
considered. The conventional dynamic formalism, obtained for the
space-time, where the motion of free particles is primordially
deterministic, seems to be unsuitable. The statistical ensemble of
stochastic (or deterministic) systems is considered to be the main
object of dynamics. \label{0b} At such a logical reloading the
statistical description becomes a component of dynamics, and
capacities of dynamics increase.\label{0e} At such an approach one
can describe deterministic, stochastic and quantum particles by
means of the uniform technique. The quantum particle is described
as a stochastic particle, i.e. without a reference to the quantum
principles. Besides, by means of this technique one can describe
classical inviscid fluid. There are four different versions of the
formalism: (1) description in Euler dynamic variables, (2)
description in Lagrange dynamic variables, (3) description in
terms of the generalized stream function, (4) description in terms
of the wave function. The uniform formalism is purely dynamic.
Even describing stochastic systems, it does not refer to
probability and probabilistic structures. In relativistic case the
uniform formalism can describe pair production and pair
annihilation.
\end{abstract}

\maketitle

\section{Introduction}

The classical mechanics and the infinitesimal calculus had been
created by Isaac Newton in the 17th century practically
simultaneously. The ordinary differential equations were the
principal tool of the classical mechanics. In the 18th and 19th
centuries the development of the classical mechanics was carried
out by means of modification of dynamic equations, when they were
applied to new dynamic systems. All this time the conception of
the event space (space-time) retained to be unchanged. In the
Newtonian conception of the event space there are two independent
invariants: space distance and time interval.

In the beginning of the 20th century Albert Einstein had
discovered, that the dynamics may be developed not only by means
of a modification of dynamic equations. The dynamics may be
developed also by a modification of the event space. A. Einstein
suggested and carried out the first two modifications of the event
space. In the first modification, known as the special relativity,
two Newtonian invariants (space and time) were replaced by one
invariant -- the space-time interval. After such a replacement one
may speak about the space-time and the space-time geometry. The
second modification of the event space was produced by A.Einstein
ten years later. This modification is known as the general
relativity. According to the general relativity the space-time
geometry may be inhomogeneous, and this nonhomogeneity depends on
the matter distribution in the space-time.

In the thirties of the 20th century it was discovered that the
free particles of small mass move stochastically. \textit{The
motion of free particles depends only on the space-time
properties.} It meant that for the explanation of the observed
stochasticity one needs the next modification of the event space.
The necessary third modification of the space-time geometry were
to look rather exotic. As far as the stochasticity was different
for the particles of different mass, the free motion of a particle
must depend on the particle mass, i.e. the particle mass is to be
geometrized. In the framework of the Riemannian geometry it was
impossible. Besides, in the framework of the classical mechanics
the particle motion is deterministic. If we want to explain the
particle motion stochasticity by the space-time properties, we are
to use such a space-time geometry, where the free particle motion
be primordially stochastic. In the framework of the Riemannian
geometry it was impossible. We did not know a geometry with such
properties.

In the beginning of the 20th century we had the alternative:
either space-time geometry with unusual exotic properties, or a
refusal from the classical mechanics. We had no adequate
space-time geometry. The alternative was resolved in favour of the
quantum mechanics, which substituted the classical mechanics of
the small mass particles. At such a substitution the principles of
the classical mechanics were replaced by the quantum principles.
Such a replacement of the classical principles by the quantum
principles was a very complicated procedure, which was produced
only for nonrelativistic phenomena.

The modification of the space-time geometry is more attractive
from logical viewpoint, than the quantum mechanics, because it
changes only space-time properties, but does not change classical
principles of dynamics, whereas the quantum mechanics revises
these principles. Unfortunately, the third modification of the
space-time geometry was impossible in the first half of the 20th
century.

The new conception of geometry, which made possible the third
modification of the space-time \cite{R91}, appeared only in the
end of the 20th century. The new conception of geometry (known as
T-geometry \cite{R90,R02,R2005}) is very simple. It supposes that
any space-time geometry is described completely by the world
function $\sigma$ \cite{S60}, and then any space-time geometry can
be obtained from the proper Euclidean geometry by means of a
deformation (replacement of the Euclidean world function $\sigma
_{\mathrm{E}}$ with the world function $\sigma $ of the geometry
in question in all definitions and relations of the proper
Euclidean geometry). The T-geometry has unusual properties. In
general, it is multivariant and nonaxiomatizable. It can be
represented in coordinateless form. Discrete T-geometry and
continuous one are described uniform.

In the new (nondegenerate) space-time geometry the particle mass
is geometrized, the free particle motion is primordially
stochastic, and the parallelism of vectors is absolute and
intransitive, in general. Besides, parameters of the space-time
depend on the quantum constant, and the statistical description of
stochastic particles motion is equivalent to the quantum
description. A use of the nondegenerate space-time geometry admits
one to return to the classical mechanics of stochastic particles,
eliminating quantum principles.

In general, the infinitesimal calculus, created for the Newtonian
event space with the deterministic particle motion, disagrees with
the space-time conception, where the free particle motion is
primordially stochastic. One needs a new mathematical tool, which
be in accordance with the nondegenerate space-time geometry.
Construction of such a tool is a very difficult problem, and we
shall not try to solve it. Instead, we \textit{take from the new
space-time geometry only the property of stochastic motion of free
particles and try to describe it in the conventional Riemannian
space-time}. In the Riemannian space-time the natural motion of
free particles is deterministic. We try to formulate the
conventional classical mechanics in such a form, where the
stochastic particle motion be natural, whereas the deterministic
particle motion be a special case of the stochastic motion, when
the stochasticity vanishes.

Note that the conventional classical mechanics considers only
deterministic particles, whose motion is described by dynamic
equations (ordinary differential equations). One considers only
some special cases of stochastic motion, referring to the
probability theory in this consideration. The main object of the
conventional classical mechanics is a single deterministic
particle $\mathcal{S}_{\mathrm{d}}$. From viewpoint of the
conventional classical mechanics the deterministic particle
(dynamic system) and the stochastic particle (stochastic system)
are conceptually different objects. The conceptual difference
consists in the fact that there are dynamic equations for the
dynamic system and there are no dynamic equations for the
stochastic system. There is not even a collective concept with
respect to concept of dynamic system and that of stochastic
system.

If information on the deterministic particle
$\mathcal{S}_{\mathrm{d}}$ is incomplete (for instance, if the
initial conditions are known
approximately), there are different versions of the particle $\mathcal{S}_{%
\mathrm{d}}$ motion. The particle motion is multiple-path. In this
case we
consider all these possible versions. One uses the statistical ensemble $%
\mathcal{E}\left[ \mathcal{S}_{\mathrm{d}}\right] $, which is the
set of
many independent particles $\mathcal{S}_{\mathrm{d}}$. Different elements $%
\mathcal{S}_{\mathrm{d}}$ of the statistical ensemble
$\mathcal{E}\left[ \mathcal{S}_{\mathrm{d}}\right] $ move
differently, and motion of all these
elements describe all possible motions of the particle $\mathcal{S}_{\mathrm{%
d}}$. The particle $\mathcal{S}_{\mathrm{d}}$ is a dynamic system,
the statistical ensemble $\mathcal{E}\left[
\mathcal{S}_{\mathrm{d}}\right] $ is also a dynamic system. It
means that there are dynamic equations for both
the particle $\mathcal{S}_{\mathrm{d}}\ $and the statistical ensemble $%
\mathcal{E}\left[ \mathcal{S}_{\mathrm{d}}\right] $. These dynamic
equations
describe the state evolution respectively of the particle $\mathcal{S}_{%
\mathrm{d}}$ and of the statistical ensemble $\mathcal{E}\left[ \mathcal{S}_{%
\mathrm{d}}\right] $.

Dynamic equations for $\mathcal{S}_{\mathrm{d}}$ and for
$\mathcal{E}\left[ \mathcal{S}_{\mathrm{d}}\right] $ are connected
between themselves. For instance, if the dynamic system
$\mathcal{S}_{\mathrm{d}}$ is a free nonrelativistic particle, the
action $\mathcal{A}_{\mathcal{S}_{\mathrm{d}}}$ for
$\mathcal{S}_{\mathrm{d}}$ has the form
\begin{equation}
\mathcal{A}_{\mathcal{S}_{\mathrm{d}}}\left[ \mathbf{x}\right] =\int \frac{m%
}{2}\mathbf{\dot{x}}^{2}dt,\qquad \mathbf{\dot{x}\equiv }\frac{d\mathbf{x}}{%
dt}  \label{a1.1}
\end{equation}%
where $\mathbf{x}=\mathbf{x}\left( t\right) =\left\{ x^{1}\left(
t\right) ,x^{2}\left( t\right) ,x^{3}\left( t\right) \right\} $,
and $m$ is the particle mass.

The action for the statistical ensemble $\mathcal{E}\left[ \mathcal{S}_{%
\mathrm{d}}\right] $ of free independent particles
$\mathcal{S}_{\mathrm{d}}$ is the sum of actions (\ref{a1.1}). It
has the form
\begin{equation}
\mathcal{A}_{\mathcal{E}\left[ \mathcal{S}_{\mathrm{d}}\right]
}\left[
\mathbf{x}\right] =\int \int\limits_{V_{\xi }}\frac{m}{2}\mathbf{\dot{x}}%
^{2}\rho _{0}\left( \mathbf{\xi }\right) dtd\mathbf{\xi },\qquad \mathbf{%
\dot{x}\equiv }\frac{d\mathbf{x}}{dt}  \label{d1.2}
\end{equation}%
where $\mathbf{x}=\mathbf{x}\left( t,\mathbf{\xi }\right) =\left\{
x^{1}\left( t,\mathbf{\xi }\right) ,x^{2}\left( t,\mathbf{\xi
}\right) ,x^{3}\left( t,\mathbf{\xi }\right) \right\} ,$\
$\mathbf{\xi =}\left\{ \xi _{1},\xi _{2},\xi _{3}\right\} $ are
variables (Lagrangian coordinates), which label elements
(particles) of the statistical ensemble. $V_{\xi }$ is the region
of variables $\mathbf{\xi }$. The quantity $\rho _{0}\left(
\mathbf{\xi }\right) $ is the weight function. The quantity%
\begin{equation}
N=\int_{V_{\xi }}\rho _{0}\left( \mathbf{\xi }\right) d\mathbf{\xi
} \label{d1.2a}
\end{equation}%
may be interpreted as the number of dynamic systems $\mathcal{S}_{\mathrm{d}%
} $, constituting the statistical ensemble.

Dynamic equations, generated by the actions (\ref{a1.1}) and
(\ref{d1.2}) are similar
\begin{equation}
m\frac{d^{2}\mathbf{x}}{dt^{2}}=0,\qquad
\mathbf{x}=\mathbf{x}\left( t\right) \label{d1.3}
\end{equation}%
\begin{equation}
\rho _{0}\left( \mathbf{\xi }\right) m\frac{d^{2}\mathbf{x}}{dt^{2}}%
=0,\qquad \mathbf{x}=\mathbf{x}\left( t,\mathbf{\xi }\right)
\label{d1.4}
\end{equation}%
The dynamic systems $\mathcal{S}_{\mathrm{d}}$ and
$\mathcal{E}\left[ \mathcal{S}_{\mathrm{d}}\right] $ are
equivalent in the sense that one can
obtain dynamic equations for $\mathcal{E}\left[ \mathcal{S}_{\mathrm{d}}%
\right] $ from dynamic equations for $\mathcal{S}_{\mathrm{d}}$.
Vice versa, one can obtain dynamic equations for
$\mathcal{S}_{\mathrm{d}}$ from dynamic equations for
$\mathcal{E}\left[ \mathcal{S}_{\mathrm{d}}\right] $.

What of dynamic systems $\mathcal{S}_{\mathrm{d}}$ and
$\mathcal{E}\left[ \mathcal{S}_{\mathrm{d}}\right] $ is primary,
and what is derivative?
Conventionally, one supposes that the dynamic system $\mathcal{S}_{\mathrm{d}%
}$ is a primary fundamental object, whereas the statistical
ensemble is considered usually as a secondary derivative object,
because it is a more complicated object, consisting of
$\mathcal{S}_{\mathrm{d}}$.

We suggest to consider the statistical ensemble $\mathcal{E}\left[ \mathcal{S%
}_{\mathrm{d}}\right] $ to be the primary object, whereas the
single dynamic system $\mathcal{S}_{\mathrm{d}}$ is considered to
be the secondary derivative object. Such an approach admits one to
construct dynamics of stochastic systems.

Indeed, the statistical ensemble $\mathcal{E}\left[ \mathcal{S}_{\mathrm{st}}%
\right] $ of independent stochastic systems
$\mathcal{S}_{\mathrm{st}}$ is a dynamic system, although
$\mathcal{S}_{\mathrm{st}}$ is a stochastic system.
It means that there exist dynamic equations for $\mathcal{E}\left[ \mathcal{S%
}_{\mathrm{st}}\right] $, although there are no dynamic equations for $%
\mathcal{S}_{\mathrm{st}}$. Explanation of this surprising fact is
as follows. When we construct the statistical ensemble of many
stochastic systems, the regular features are accumulated, whereas
random features are compensated. As a result, if the number of
stochastic systems tends to infinity, we obtain the system, having
only regular characteristics. In other words, we obtain a dynamic
system.

Mathematically it looks as follows. We add some terms to the action (\ref%
{d1.2}). These terms are chosen in such a way, to describe the
quantum stochasticity, generated by the properties of the
space-time geometry. The supposed method of taking into account of
this stochasticity leads to the quantum description, which has
been well investigated. The action is written
in the form%
\begin{equation}
\mathcal{A}_{\mathcal{E}\left[ \mathcal{S}_{\mathrm{st}}\right]
}\left[
\mathbf{x},\mathbf{u}\right] =\int \int\limits_{V_{\xi }}\left\{ \frac{m}{2}%
\mathbf{\dot{x}}^{2}+\frac{m}{2}\mathbf{u}^{2}-\frac{\hbar }{2}\mathbf{%
\nabla u}\right\} \rho _{0}\left( \mathbf{\xi }\right) dtd\mathbf{\xi }%
,\qquad \mathbf{\dot{x}\equiv }\frac{d\mathbf{x}}{dt}
\label{d1.5}
\end{equation}%
The variable $\mathbf{x}=\mathbf{x}\left( t,\mathbf{\xi }\right) $
describes
the regular component of the particle motion. The variable $\mathbf{u}=%
\mathbf{u}\left( t,\mathbf{x}\right) $ describes the mean value of
the stochastic velocity component, $\hbar $ is the quantum
constant. The second term in (\ref{d1.5}) describes the kinetic
energy of the stochastic velocity component. The third term
describes interaction between the stochastic
component $\mathbf{u}\left( t,\mathbf{x}\right) $ and the regular component $%
\mathbf{\dot{x}}\left( t,\mathbf{\xi }\right) $. The operator
\begin{equation}
\mathbf{\nabla =}\left\{ \frac{\partial }{\partial x^{1}},\frac{\partial }{%
\partial x^{2}},\frac{\partial }{\partial x^{3}}\right\}  \label{d1.5a}
\end{equation}%
is defined in the space of coordinates $\mathbf{x}$. Dynamic
equations for the dynamic system $\mathcal{E}\left[
\mathcal{S}_{\mathrm{st}}\right] $ are obtained as a result of
variation of the action (\ref{d1.5}) with respect to dynamic
variables $\mathbf{x}$ and $\mathbf{u}$.

To obtain the action functional for $\mathcal{S}_{\mathrm{st}}$
from the
action (\ref{d1.5}) for $\mathcal{E}\left[ \mathcal{S}_{\mathrm{st}}\right] $%
, we should omit integration over $\mathbf{\xi }$ in (\ref{d1.5}),
as it
follows from comparison of (\ref{d1.2}) and (\ref{a1.1}). We obtain%
\begin{equation}
\mathcal{A}_{\mathcal{S}_{\mathrm{st}}}\left[
\mathbf{x},\mathbf{u}\right]
=\int \left\{ \frac{m}{2}\mathbf{\dot{x}}^{2}+\frac{m}{2}\mathbf{u}^{2}-%
\frac{\hbar }{2}\mathbf{\nabla u}\right\} dt,\qquad \mathbf{\dot{x}\equiv }%
\frac{d\mathbf{x}}{dt}  \label{d1.6}
\end{equation}%
where $\mathbf{x}=\mathbf{x}\left( t\right) $ and $\mathbf{u}=\mathbf{u}%
\left( t,\mathbf{x}\right) $ are dependent dynamic variables. The
action functional (\ref{d1.6}) is not well defined (for $\hbar
\neq 0$), because the operator $\mathbf{\nabla }$ is defined in
some 3-dimensional vicinity of point $\mathbf{x}$, but not at the
point $\mathbf{x}$ itself. As far as the action functional
(\ref{d1.6}) is not well defined, one cannot obtain dynamic
equations for $\mathcal{S}_{\mathrm{st}}$. By definition it means
that the particle $\mathcal{S}_{\mathrm{st}}$ is stochastic.
Setting $\hbar
=0$ in (\ref{d1.6}), we transform the action (\ref{d1.6}) into the action (%
\ref{a1.1}), because in this case $\mathbf{u}=0$ in virtue of
dynamic equations.

\label{1beg} Dependence of the mean stochastic velocity
$\mathbf{u}$ on dependent variable $\mathbf{x}$, describing a
regular component of the motion, and appearance of $\mathbf{\nabla
u}$ in the action functional for the statistical ensemble
$\mathcal{E}\left[ \mathcal{S}_{\mathrm{st}}\right] $ are a formal
sign of the particle stochasticity.

The quantum constant $\hbar $ has been introduced in the action (\ref{d1.5}%
), in order the description by means of the action (\ref{d1.5}) be
equivalent to the quantum description by means of the
Schr\"{o}dinger equation \cite{R99}. If we substitute the term
$-\hbar \mathbf{\nabla u/}2$ by some function $f\left(
\mathbf{\nabla u}\right) $, we obtain statistical description of
other stochastic system with other form of stochasticity, which
does not coincide with the quantum stochasticity. In other words,
the form of the last term in (\ref{d1.5}) describes the type of
the stochasticity.

Although we cannot investigate the stochastic particle $\mathcal{S}_{\mathrm{%
st}}$, we can describe and investigate the statistical ensemble $\mathcal{E}%
\left[ \mathcal{S}_{\mathrm{st}}\right] $ of stochastic particles $\mathcal{S%
}_{\mathrm{st}}$, because $\mathcal{E}\left[
\mathcal{S}_{\mathrm{st}}\right] $ is well defined dynamic system
(\ref{d1.5}). Investigation of the statistical ensemble
$\mathcal{E}\left[ \mathcal{S}_{\mathrm{st}}\right] $ admits one
to investigate some average characteristics of the stochastic
particle $\mathcal{S}_{\mathrm{st}}$. Information on $\mathcal{S}_{\mathrm{st%
}}$, obtained at investigation of $\mathcal{E}\left[ \mathcal{S}_{\mathrm{st}%
}\right] $, is not a full information. One may obtain the mean
velocity of the stochastic particle, the mean trajectories, the
mean energy and some other average characteristics. However, one
cannot obtain the velocity distribution and other more detailed
characteristics of the stochastic particle. To obtain such
detailed characteristics, one needs to use additional information
on the stochastic particle properties (investigation
of the statistical ensemble $\mathcal{E}\left[ \mathcal{S}_{\mathrm{st}}%
\right] $ is insufficient for this goal). Nevertheless, the
information, which is obtained from investigation of the
statistical ensemble, appears to be valuable in many cases. For
instance, one can show \cite{R99} that by a proper change of
variables the action (\ref{d1.5}) is reduced to the action for the
Schr\"{o}dinger particle, i.e. to the action for the dynamic
system, described by the Schr\"{o}dinger equation.

Thus, there is a general approach to a description of stochastic
particles, when the deterministic particle is considered to be a
special case of stochastic particle (with vanishing
stochasticity). To realize this
approach, we are to\textit{\ consider the statistical ensemble }$\mathcal{E}%
\left[ \mathcal{S}\right] $ \textit{as the primary object (basic
object) of dynamics, }whereas the single system $\mathcal{S}$ is
considered to be a derivative object of dynamics. Realizing this
approach, it is useful to introduce a collective concept with
respect to concept of dynamic system and that of stochastic
system. We shall use the term "physical system". We shall speak
about the statistical ensemble $\mathcal{E}\left[
\mathcal{S}\right] $
of physical systems $\mathcal{S}$, and it is of no importance, whether $%
\mathcal{S}$ is the dynamic system, or the stochastic one. To
stress that the dynamic system and the stochastic system are
special cases of the physical system, we shall use the term
"deterministic physical system" instead of the term "dynamic
system" and the term "stochastic physical system" instead of the
term "stochastic system". The fact that we can obtain dynamic
equations for the dynamic system $\mathcal{S}$ and cannot obtain
them for the stochastic system $\mathcal{S}$, will be considered
as a
special property of the statistical ensemble $\mathcal{E}\left[ \mathcal{S}%
\right] $. There is a formal criterion, which admits one to
determine, whether the physical systems $\mathcal{S}$,
constituting the statistical ensemble $\mathcal{E}\left[
\mathcal{S}\right] $, are stochastic systems.
(Dynamic equations for statistical ensemble $\mathcal{E}\left[ \mathcal{S}_{%
\mathrm{d}}\right] $ can be reduced to the system of ordinary
differential equations, whereas for $\mathcal{E}\left[
\mathcal{S}_{\mathrm{st}}\right] $ such a reduction is
impossible). The fact, that we cannot obtain a description
(dynamic equations) for the single stochastic particle, is of no
importance, because the basic object of dynamics is a statistical
ensemble, and we can always obtain the description of the
statistical ensemble.

Let us return to the action (\ref{d1.5}) and obtain dynamic
equations for the statistical ensemble $\mathcal{E}\left[
\mathcal{S}_{\mathrm{st}}\right] $ of physical systems
$\mathcal{S}_{\mathrm{st}}$. Variation of (\ref{d1.5})
with respect to $\mathbf{u}$ gives%
\begin{eqnarray*}
\delta \mathcal{A}_{\mathcal{E}\left[ \mathcal{S}_{\mathrm{st}}\right] }%
\left[ \mathbf{x},\mathbf{u}\right] &=&\int \int\limits_{V_{\xi }}\left\{ m%
\mathbf{u}\delta \mathbf{u}-\frac{\hbar }{2}\mathbf{\nabla }\delta \mathbf{u}%
\right\} \rho _{0}\left( \mathbf{\xi }\right) dtd\mathbf{\xi } \\
&=&\int \int\limits_{V_{\mathbf{x}}}\left\{ m\mathbf{u}\delta \mathbf{u}-%
\frac{\hbar }{2}\mathbf{\nabla }\delta \mathbf{u}\right\} \rho
_{0}\left( \mathbf{\xi }\right) \frac{\partial \left( \xi _{1},\xi
_{2},\xi _{3}\right)
}{\partial \left( x^{1},x^{2},x^{3}\right) }dtd\mathbf{x} \\
&=&\int \int\limits_{V_{\mathbf{x}}}\delta \mathbf{u}\left\{ m\mathbf{u}%
\rho +\frac{\hbar }{2}\mathbf{\nabla }\rho \right\}
dtd\mathbf{x-}\int \oint \frac{\hbar }{2}\rho \delta
\mathbf{u}dtd\mathbf{S}
\end{eqnarray*}%
where%
\begin{equation}
\rho =\rho _{0}\left( \mathbf{\xi }\right) \frac{\partial \left(
\xi
_{1},\xi _{2},\xi _{3}\right) }{\partial \left( x^{1},x^{2},x^{3}\right) }%
=\rho _{0}\left( \mathbf{\xi }\right) \left( \frac{\partial \left(
x^{1},x^{2},x^{3}\right) }{\partial \left( \xi _{1},\xi _{2},\xi
_{3}\right) }\right) ^{-1}  \label{d1.7a}
\end{equation}%
We obtain the following dynamic equation%
\begin{equation}
m\rho \mathbf{u}+\frac{\hbar }{2}\mathbf{\nabla }\rho =0,\qquad
\label{d1.7}
\end{equation}%
Variation of (\ref{d1.5}) with respect to $\mathbf{x}$ gives%
\begin{equation}
m\frac{d^{2}\mathbf{x}}{dt^{2}}=\mathbf{\nabla }\left( \frac{m}{2}\mathbf{u}%
^{2}-\frac{\hbar }{2}\mathbf{\nabla u}\right)  \label{d1.8}
\end{equation}%
Here $d/dt$ means the substantial derivative with respect to time $t$%
\[
\frac{dF}{dt}\equiv \frac{\partial \left( F,\xi _{1},\xi _{2},\xi
_{3}\right) }{\partial \left( t,\xi _{1},\xi _{2},\xi _{3}\right)
}
\]

Note that without a loss of generality we may set $\rho _{0}\left( \mathbf{%
\xi }\right) =1$, because by means of change of variables
\begin{equation}
\tilde{\xi}_{1}=\int \rho _{0}\left( \mathbf{\xi }\right) d\xi
_{1},\qquad \tilde{\xi}_{2}=\xi _{2},\qquad \tilde{\xi}_{3}=\xi
_{3}  \label{d1.8b}
\end{equation}
we obtain%
\begin{equation}
\rho _{0}\left( \mathbf{\xi }\right) \frac{\partial \left( \xi
_{1},\xi
_{2},\xi _{3}\right) }{\partial \left( x^{1},x^{2},x^{3}\right) }=\frac{%
\partial \left( \int \rho _{0}\left( \mathbf{\xi }\right) d\xi _{1},\xi
_{2},\xi _{3}\right) }{\partial \left( x^{1},x^{2},x^{3}\right) }=\frac{%
\partial \left( \tilde{\xi}_{1},\tilde{\xi}_{2},\tilde{\xi}_{3}\right) }{%
\partial \left( x^{1},x^{2},x^{3}\right) }  \label{d1.8a}
\end{equation}

Resolving (\ref{d1.7}) with respect to $\mathbf{u}$, we obtain the equation%
\begin{equation}
\mathbf{u}=-\frac{\hbar }{2m}\mathbf{\nabla }\ln \rho ,
\label{d1.9}
\end{equation}%
which reminds the expression for the mean velocity of the Brownian
particle with the diffusion coefficient $D=\hbar /2m$.

Eliminating the velocity $\mathbf{u}$ from dynamic equations
(\ref{d1.8}) and (\ref{d1.9}), we obtain the dynamic equations of
the hydrodynamic type
for the mean motion of the stochastic particle $\mathcal{S}_{\mathrm{st}}$%
\begin{equation}
m\frac{d^{2}\mathbf{x}}{dt^{2}}=-\mathbf{\nabla }U_{\mathrm{B}},\qquad U_{%
\mathrm{B}}=U\left( \rho ,\mathbf{\nabla }\rho ,\mathbf{\nabla
}^{2}\rho
\right) =\frac{\hbar ^{2}}{8m}\frac{\left( \mathbf{\nabla }\rho \right) ^{2}%
}{\rho ^{2}}-\frac{\hbar ^{2}}{4m}\frac{\mathbf{\nabla }^{2}\rho
}{\rho } \label{d1.10}
\end{equation}%
Here $\rho $ is considered to be function of $t,\mathbf{x}$, and $\mathbf{%
\nabla }$ is the gradient in the space of coordinates
$\mathbf{x}$.

If
\begin{equation}
\frac{\partial \left( \xi _{1},\xi _{2},\xi _{3}\right) }{\partial
\left( x^{1},x^{2},x^{3}\right) }\neq 0  \label{d1.10a}
\end{equation}%
and the relations $\mathbf{x}=\mathbf{x}\left( t,\mathbf{\xi
}\right) $ can
be resolved with respect to variables $\mathbf{\xi }$ in the form $\mathbf{%
\xi }=x\left( t,\mathbf{x}\right) $, dynamic equations
(\ref{d1.10}) can be rewritten in the Euler form.

Using the relation (\ref{d1.7a}), one can rewrite the designation
\begin{equation}
\mathbf{v=v}\left( t,\mathbf{x}\right) =\frac{d\mathbf{x}}{dt}=\frac{%
\partial \left( \mathbf{x},\xi _{1},\xi _{2},\xi _{3}\right) }{\partial
\left( t,\xi _{1},\xi _{2},\xi _{3}\right) }=\frac{\partial \left( \mathbf{x}%
,\xi _{1},\xi _{2},\xi _{3}\right) }{\partial \left(
t,x^{1},x^{2},x^{3}\right) }\frac{\partial \left(
t,x^{1},x^{2},x^{3}\right) }{\partial \left( t,\xi _{1},\xi
_{2},\xi _{3}\right) }  \label{d1.11}
\end{equation}%
in the form%
\begin{equation}
\rho \mathbf{v=}\rho \mathbf{v}\left( t,\mathbf{x}\right) =\rho \frac{d%
\mathbf{x}}{dt}=\rho _{0}\left( \xi \right) \frac{\partial \left( \mathbf{x}%
,\xi _{1},\xi _{2},\xi _{3}\right) }{\partial \left(
t,x^{1},x^{2},x^{3}\right) }  \label{d1.12}
\end{equation}

Substituting (\ref{d1.11}) in (\ref{d1.10}), we obtain the Euler
form of
hydrodynamic equations%
\begin{equation}
\frac{d\mathbf{v}}{dt}\equiv \frac{\partial \mathbf{v}}{\partial
t}+\left(
\mathbf{v\nabla }\right) \mathbf{v}=-\frac{1}{m}\mathbf{\nabla }U_{\mathrm{B}%
}  \label{d1.14}
\end{equation}%
Instead of the definition (\ref{d1.7a}) the quantity $\rho $ in (\ref{d1.10}%
) and (\ref{d1.14}) is considered as $\rho =\rho \left( t,\mathbf{x}\right) $%
, which is determined from the continuity equation%
\begin{equation}
\frac{\partial \rho }{\partial t}+\mathbf{\nabla }\left( \rho \mathbf{v}%
\right) =0  \label{d1.15}
\end{equation}%
The continuity equation is fulfilled identically in force of relations (\ref%
{d1.7a}) and (\ref{d1.12}).

Two equations (\ref{d1.14}), (\ref{d1.15}) describe evolution of
the statistical ensemble of stochastic particles. Character of
stochasticity is determined by the Bohm potential $U_{\mathrm{B}}$
\cite{B52}, defined by the relation (\ref{d1.10}). Any reference
to the stochastic velocity distribution or to some other
probability distribution is absent. Influence of this distribution
on the mean motion of the particles is described by the form of
interaction (\ref{d1.10}). The situation reminds the case of the
gas dynamics, where the action of the Maxwell velocity
distribution on the gas motion is described by the internal gas
energy. Of course, such a description is not comprehensive,
however, it is sufficient for a description of the mean motion of
the stochastic particle. As a result we obtain a \textit{purely
dynamic description} of the stochastic particle motion.

Equations for the ideal fluid may be described in terms of the
wave function \cite{R99}. Irrotational flow of the fluid
(\ref{d1.10}) is described by the Schr\"{o}dinger equation for the
free quantum particle. It means:

\begin{enumerate}
\item The statistical ensemble of free quantum nonrelativistic
particles may be considered to be a statistical ensemble of
stochastic particles, which is described by the action
(\ref{d1.5}).

\item The wave function is simply a method of the ideal fluid
description, but not a specific quantum object, defined by means
of enigmatic quantum principles.
\cutpage 

\setcounter{page}{11}

\noindent
 \item The quantum particles are stochastic particles,
which may be described in terms of dynamics of physical systems,
where the basic object is the statistical ensemble.
\end{enumerate}
 Thus, the dynamics of physical systems admits one to
describe quantum effects without a reference to quantum
principles, because the quantum particles, as well as stochastic
ones are objects of classical dynamics of physical systems.
Description of stochastic and quantum particles is the problem of
the classical dynamics, where the \textit{basic object of dynamics
is the statistical ensemble}.

\section{Dynamics of arbitrary physical systems}

The action (\ref{d1.5}) for the statistical ensemble of free
nonrelativistic stochastic particles may be easily generalized to
the case of arbitrary stochastic systems. Let
$\mathcal{S}_{\mathrm{d}}$ be a deterministic physical system
having the finite number of the freedom degrees. The state
of $\mathcal{S}_{\mathrm{d}}$ is described by the generalized coordinates $%
\mathbf{x}=\left\{ x^{1},x^{2},...x^{n}\right\} $. The action has
the form
\begin{equation}
\;\mathcal{A}_{\mathcal{S}_{\mathrm{d}}}\left[ \mathbf{x}\right] =\int L_{%
\mathrm{d}}\left( t,\mathbf{x},\mathbf{\dot{x},}P\right) dt,\qquad \mathbf{%
\dot{x}\equiv }\frac{d\mathbf{x}}{dt}  \label{d2.1}
\end{equation}%
where $\mathbf{x}=\mathbf{x}\left( t\right) $ and $P$ are some
parameters of the system (for instance, masses, charges, etc.)

Statistical ensemble $\mathcal{E}\left[
\mathcal{S}_{\mathrm{d}}\right] $ of dynamic systems
$\mathcal{S}_{\mathrm{d}}$ is described by the action
\begin{equation}
\mathcal{A}_{\mathcal{E}\left[ \mathcal{S}_{\mathrm{d}}\right]
}\left[ \mathbf{x}\right] =\int \int\limits_{V_{\mathbf{\xi
}}}L_{\mathrm{d}}\left( t,\mathbf{x},\mathbf{\dot{x},}P\right)
\rho _{0}\left( \mathbf{\xi }\right) dtd^{n}\xi ,\qquad
\mathbf{\dot{x}\equiv }\frac{d\mathbf{x}}{dt} \label{d2.2}
\end{equation}%
where $\mathbf{x}=\mathbf{x}\left( t,\mathbf{\xi }\right) =\left\{
x^{1}\left( t,\mathbf{\xi }\right) ,x^{2}\left( t,\mathbf{\xi
}\right)
,...x^{n}\left( t,\mathbf{\xi }\right) \right\} $. The variables $\mathbf{%
\xi }=\left\{ \xi _{1},\xi _{2},...\xi _{n}\right\} $ label elements $%
\mathcal{S}_{\mathrm{d}}$ of the statistical ensemble. The
quantity $\rho _{0}\left( \mathbf{\xi }\right) $ is the weight
function. The number $k$ of the labelling variables is chosen to
be equal to the number $n$ of generalized coordinates, in order
one can to pass to the independent
variables $t,\mathbf{x}$\textbf{, }resolving relations $\mathbf{x}=\mathbf{x}%
\left( t,\mathbf{\xi }\right) $ in the form $\mathbf{\xi }=\mathbf{\xi }%
\left( t,\mathbf{x}\right) $. If we are not going to pass to
independent variables $t,\mathbf{x}$\textbf{, } the integer number
$k>0$ may be chosen arbitrary.

If some disturbing agent influences on the deterministic system $\mathcal{S}%
_{\mathrm{d}}$, it turns into the stochastic system $\mathcal{S}_{\mathrm{st}%
}$ and the action (\ref{d2.2}) turns into the action $\mathcal{A}_{\mathcal{E%
}\left[ \mathcal{S}_{\mathrm{st}}\right] }$%
\begin{equation}
\mathcal{A}_{\mathcal{E}\left[ \mathcal{S}_{\mathrm{st}}\right]
}\left[
\mathbf{x,}u\right] =\int \int\limits_{V_{\mathbf{\xi }}}L\left( t,\mathbf{x%
},\mathbf{\dot{x}},P_{\mathrm{eff}}\left( u\right) \right) \rho
_{0}\left(
\mathbf{\xi }\right) dtd^{n}\xi ,\qquad \mathbf{\dot{x}\equiv }\frac{d%
\mathbf{x}}{dt}  \label{d2.3}
\end{equation}%
where $\mathbf{x}=\mathbf{x}\left( t,\mathbf{\xi }\right) $ and $%
u^{k}=\left\{ u^{k}\left( t,\mathbf{x}\right) \right\} ,$
$k=0,1,...n,$ are dependent variables. The new dependent variables
$u^{k}$ describe the mean
value of the stochastic component of the generalized velocity $\mathbf{\dot{x%
}}$. It is supposed, that the disturbing agent changes the values
of the
parameters of dynamic system $S_{\mathrm{d}}$. The Lagrangian $L\left( t,%
\mathbf{x},\mathbf{\dot{x}},P_{\mathrm{eff}}\left( u\right)
\right) $ for
the statistical ensemble of the corresponding stochastic system $\mathcal{S}%
_{\mathrm{st}}$ is obtained from the Lagrangian $L_{\mathrm{d}}\left( t,%
\mathbf{x},\mathbf{\dot{x}},P\right) $ for the statistical
ensemble of the
dynamic system $\mathcal{S}_{\mathrm{d}}$ by means of the replacement \cite%
{R2002}%
\begin{equation}
P\rightarrow P_{\mathrm{eff}}\left( u\right)  \label{d2.5}
\end{equation}%
in the expression (\ref{d2.2}). Passing to description of stochastic system $%
\mathcal{S}_{\mathrm{st}}$, we \textit{do not introduce any
probabilistic structures, and the descriptions remains to be
purely dynamic}. Character of stochasticity is determined by the
form of the change (\ref{d2.5}).

In the case, when the dynamic system $\mathcal{S}_{\mathrm{d}}$ is
the free uncharged relativistic particle, the only parameter $P$
is the particle mass $m$. If the stochastic agent is the
distortion of the space-time geometry,
the replacement (\ref{d2.5}) has the form%
\begin{equation}
m\rightarrow m_{\mathrm{eff}}=\sqrt{m^{2}+\frac{\hbar
^{2}}{c^{2}}\left( g_{kl}\kappa ^{k}\kappa ^{l}+\partial
_{k}\kappa ^{k}\right) }  \label{d2.6}
\end{equation}%
where $c$ is the speed of the light, $g_{kl}=$diag$\{
c^{2},$$-1,-1,-1\} $ is the metric tensor,
\begin{equation}
\kappa ^{k}=\frac{m}{\hbar }u^{k},\qquad k=0,1,2,3  \label{d2.7}
\end{equation}%
and $u^{k}\left( t,\mathbf{x}\right) =u^{k}\left( x\right) $ is
the mean value of the stochastic component of the particle
4-velocity. Here and later
on there is a summation over repeating indices: $0-3$ for Latin indices and $%
1-3$ for Greek ones.

In the relativistic case the action for the statistical ensemble (\ref{d2.3}%
) has the form%
\begin{equation}
\mathcal{A}_{\mathcal{E}\left[ \mathcal{S}_{\mathrm{st}}\right] }\left[ x%
\mathbf{,}\kappa \right] =-\int \int\limits_{V_{\mathbf{\xi }}}mcK\sqrt{%
g_{ik}\dot{x}^{i}\dot{x}^{k}}\rho _{0}\left( \mathbf{\xi }\right) d\tau d%
\mathbf{\xi },\qquad \mathbf{\dot{x}\equiv
}\frac{d\mathbf{x}}{d\tau } \label{d2.8}
\end{equation}%
\begin{equation}
K=\sqrt{1+\lambda ^{2}\left( g_{kl}\kappa ^{k}\kappa ^{l}+\partial
_{k}\kappa ^{k}\right) },\qquad \lambda =\frac{\hbar }{mc}
\label{d2.9}
\end{equation}%
where $x=\left\{ x^{k}\right\} =\left\{ x^{k}\left( \tau ,\mathbf{\xi }%
\right) \right\} ,$ $k=0,1,2,3$. The quantity $g_{kl}=$diag$\{
c^{2},-1,$ $-1,-1\} $ is the metric tensor. The independent variables $%
\mathbf{\xi }=\left\{ \xi _{1},\xi _{2},\xi _{3}\right\} $ label
the particles of the statistical ensemble. The dependent variables
$\kappa ^{k}=\kappa ^{k}\left( x\right) $, $k=0,1,2,3$ form some
force field,
connected with the stochastic component of the particle 4-velocity, and $%
\lambda $ is the Compton wave length of the particle.

In the nonrelativistic approximation, one may neglect the temporal
component
$\kappa ^{0}=\frac{m}{\hbar }u^{0}$ with respect to the spatial one $\mathbf{%
\kappa }=\frac{m}{\hbar }\mathbf{u}.$ Setting $\tau =t$ $=x^{0}$ in (\ref%
{d2.8}), (\ref{d2.9}) we obtain instead of (\ref{d2.8})
\begin{equation}
\mathcal{A}_{\mathcal{E}\left[ \mathcal{S}_{\mathrm{st}}\right]
}\left[ \mathbf{x},\mathbf{u}\right] =\int
\int\limits_{V_{\mathbf{\xi }}}\left\{
-mc^{2}+\frac{m}{2}\mathbf{\dot{x}}^{2}+\frac{m}{2}\mathbf{u}^{2}-\frac{%
\hbar }{2}\mathbf{\nabla u}\right\} \rho _{0}\left( \mathbf{\xi }\right) dtd%
\mathbf{\xi },\qquad \mathbf{\dot{x}\equiv }\frac{d\mathbf{x}}{dt}
\label{d2.10}
\end{equation}

The action (\ref{d2.10}) coincides with the action (\ref{d1.5})
except for the first term, which does not contribute to dynamic
equations.

In the relativistic case, varying (\ref{d2.8}) with respect to $\kappa ^{i}$%
, we obtain the dynamic equations%
\begin{equation}
\frac{\delta \mathcal{A}}{\delta \kappa ^{i}}=-\lambda ^{2}\frac{mc\sqrt{%
g_{ik}\dot{x}^{i}\dot{x}^{k}}\rho _{0}\left( \mathbf{\xi }\right) }{K}%
g_{ik}\kappa ^{k}+\lambda ^{2}\partial _{i}\frac{mc\sqrt{g_{ik}\dot{x}^{i}%
\dot{x}^{k}}\rho _{0}\left( \mathbf{\xi }\right) }{2K}=0
\label{d2.10a}
\end{equation}%
These equations are integrated in the form%
\begin{equation}
\kappa =\frac{1}{2}\log
\frac{C\sqrt{g_{ik}\dot{x}^{i}\dot{x}^{k}}\rho _{0}\left(
\mathbf{\xi }\right) }{K},\qquad C=\text{const}  \label{d2.12}
\end{equation}%
where the quantity $\kappa $ is a potential for the field $\kappa ^{k}$%
\begin{equation}
\partial _{k}\kappa =g_{kl}\kappa ^{l},\qquad k=0,1,2,3  \label{d2.12a}
\end{equation}%
The dynamic equation (\ref{d2.12}) may be rewritten in the form%
\begin{equation}
e^{2\kappa }=C\frac{\sqrt{g_{ik}\dot{x}^{i}\dot{x}^{k}}\rho
_{0}\left( \mathbf{\xi }\right) }{\sqrt{1+\lambda
^{2}g^{ls}e^{-\kappa }\partial _{k}\partial ^{k}e^{\kappa
}}},\qquad C=\text{const}  \label{d2.11}
\end{equation}%
which is an analog of nonrelativistic dynamic equation
(\ref{d1.9}).

The principal difference between the nonrelativistic description (\ref{d1.9}%
) and the relativistic description (\ref{d2.11}) is as follows.
The nonrelativistic equation (\ref{d1.9}) does not contain
temporal derivatives, and the field $\mathbf{u}$ is determined
uniquely by its source (the particle density $\rho )$. The
relativistic equation (\ref{d2.11}) contains temporal derivatives,
and the $\kappa $-field $u^{k}=\hbar \kappa ^{k}/m$ can exist
without its source. The relativistic $\kappa $-field $u^{k}=\hbar
\kappa ^{k}/m$ can escape from its source. Besides, the $\kappa
$-field
changes the effective particle mass, as one can see from the relations (\ref%
{d2.6}) or (\ref{d2.8}), (\ref{d2.9}). If $\mathbf{\kappa }^{2}$
is large enough, or $\partial _{k}\kappa ^{k}<0$ and $\left\vert
\partial _{k}\kappa ^{k}\right\vert $ is large enough, the
effective particle mass may be imaginary. In this case the mean
world line may turn in the time direction, and this turn may
appear to be connected with the pair production, or with the pair
annihilation \cite{R2002}.

In the nonrelativistic case the mean stochastic velocity
$\mathbf{u\ }$may be eliminated and replaced by its source (the
particle density $\rho $). In the relativistic case the $\kappa
$-field has in addition its own degrees of freedom, which cannot
be eliminated, replacing the $\kappa $-field by its source. The
$\kappa $-field can travel from one space-time region to another.

The uniform formalism of dynamics (with the statistical ensemble
as a basic object of dynamics) admits one to describe such a
physical phenomena, which cannot be described in the framework of
the conventional dynamic formalism, when the basic object is a
dynamic system. In particular, one can describe the pair
production effect, which cannot been described in the framework of
the classical relativistic mechanics, as well as in the framework
of the nonrelativistic quantum mechanics.

\section{Statistical average physical system}

Let $\mathcal{S}$ be a physical system and $\mathcal{E}\left[ N,\mathcal{S}%
\right] $ be the statistical ensemble, consisting of $N$
$(N\rightarrow \infty )$ physical systems $\mathcal{S}$. Let
$\mathcal{A}_{\mathcal{E}\left[ N,\mathcal{S}\right] }$ be the
action functional for the statistical ensemble $\mathcal{E}\left[
N,\mathcal{S}\right] $. As far as the
statistical ensemble consists of identical independent systems, its action $%
\mathcal{A}_{\mathcal{E}\left[ N,\mathcal{S}\right] } $ has the
property
\begin{equation}
\mathcal{A}_{\mathcal{E}\left[ aN,\mathcal{S}\right] }=a\mathcal{A}_{%
\mathcal{E}\left[ N,\mathcal{S}\right] },\text{\qquad }a>0,\qquad a=\text{%
const,\qquad }N,aN\gg 1  \label{d3.1}
\end{equation}

Basing on this property, one may introduce such a statistical ensemble $%
\left\langle \mathcal{S}\right\rangle $, whose action $\mathcal{A}%
_{\left\langle \mathcal{S}\right\rangle }$ has the form
\begin{equation}
\mathcal{A}_{\left\langle \mathcal{S}\right\rangle
}=\lim_{N\rightarrow \infty
}\frac{1}{N}\mathcal{A}_{\mathcal{E}\left[ N,\mathcal{S}\right] }
\label{d3.2}
\end{equation}%
The deterministic physical system, whose action has the form
(\ref{d3.2}), will be referred to as the \textit{statistical
average system} $\left\langle
\mathcal{S}\right\rangle $. The physical system $\left\langle \mathcal{S}%
\right\rangle $ is a dynamic system, because it is deterministic
and has the action $\mathcal{A}_{\left\langle
\mathcal{S}\right\rangle }$. It is the
average system, because its action $\mathcal{A}_{\left\langle \mathcal{S}%
\right\rangle }$ is the mean action for any system $\mathcal{S}$
of the statistical ensemble $\mathcal{E}\left[
N,\mathcal{S}\right] $. According to definition (\ref{d3.2}) the
system $\left\langle \mathcal{S}\right\rangle $ is the statistical
ensemble $\mathcal{E}\left[ N,\mathcal{S}\right] $, normalized to
one system. In accordance with the property (\ref{d3.1}) the
definition (\ref{d3.2}) of the action $\mathcal{A}_{\left\langle \mathcal{S}%
\right\rangle }$ is invariant with respect to transformation
\begin{equation}
N\rightarrow aN,\text{\qquad }a>0,\qquad a=\text{const}
\label{d3.3}
\end{equation}

Formally the statistical average system $\left\langle \mathcal{S}%
\right\rangle $ may be considered as a statistical ensemble
consisting of one system $\mathcal{S}$. Nevertheless, according to
(\ref{d3.2}) the statistical average system $\left\langle
\mathcal{S}\right\rangle $ has
statistical properties, because the action for $\left\langle \mathcal{S}%
\right\rangle \ $ is an action, constructed of the action for the
statistical ensemble $\mathcal{E}\left[ N,\mathcal{S}\right] $
with very large number $N$ of elements ($N\rightarrow \infty $).

Being a statistical ensemble, the statistical average system
$\left\langle
\mathcal{S}\right\rangle $ has some properties of the individual system $%
\mathcal{S}$. In particular, the energy $E$, the momentum
$\mathbf{p}$ and other additive quantities of $\left\langle
\mathcal{S}\right\rangle $ coincide respectively with the mean
energy $\left\langle E\right\rangle $, the mean momentum
$\left\langle \mathbf{p}\right\rangle $ and mean values of other
additive quantities of the single system $\mathcal{S}$. In other
words, in some aspects the statistical average system $\left\langle \mathcal{%
S}\right\rangle $ is perceived as a single system $\mathcal{S}$.
On the
other hand, the statistical average system $\left\langle \mathcal{S}%
\right\rangle $ does not coincide with $\mathcal{S}$, even if the
single system $\mathcal{S}$ is a deterministic physical system.

Let, for instance, the single deterministic system $\mathcal{S}$
have $n$ degrees of freedom. Let in the definition (\ref{d3.2})
the number $N$ of elements of the statistical ensemble
$\mathcal{E}\left[ N,\mathcal{S}\right]
$ be very large, but finite. In this case the statistical average system $%
\left\langle \mathcal{S}\right\rangle $ has $nN$ degrees of
freedom. The statistical average system $\left\langle
\mathcal{S}\right\rangle $ may have alternative properties of the
single system $\mathcal{S}$ simultaneously. For instance, let
$\mathcal{S}$ be a single particle in the two-slit experiment. The
individual particle $\mathcal{S}$ may pass only through one of two
open slits, whereas the statistical average particle $\left\langle
\mathcal{S}\right\rangle $ may pass through both slits
simultaneously. The state of $nk$ freedom degrees of $\left\langle
\mathcal{S}\right\rangle $ correspond to the passage through one
slit, whereas the state of $n\left( N-k\right) $ freedom degrees
of $\left\langle \mathcal{S}\right\rangle $ correspond to the
passage through another slit.

\section{Methods of the statistical ensemble description}

We shall consider four different methods of the statistical
ensemble description: (1) description in Lagrangian coordinates,
(2) description in Eulerian coordinates, (3) Hamilton-Jacobi
description, (4) description in terms of the wave function. We
demonstrate application of these methods in the example of
nonrelativistic stochastic (quantum) particle, moving in the given
external potential $V\left( \mathbf{x}\right) $. In this case the
action (\ref{d1.5}) takes the form%
\begin{equation}
\mathcal{A}_{\mathcal{E}\left[ \mathcal{S}_{\mathrm{st}}\right]
}\left[
\mathbf{x},\mathbf{u}\right] =\int \int\limits_{V_{x}}\left\{ \frac{m}{2}%
\mathbf{\dot{x}}^{2}-V\left( \mathbf{x}\right) +\frac{m}{2}\mathbf{u}^{2}-%
\frac{\hbar }{2}\mathbf{\nabla u}\right\} \rho _{0}\left( \mathbf{\xi }%
\right) dtd\mathbf{\xi },\qquad \mathbf{\dot{x}\equiv
}\frac{d\mathbf{x}}{dt} \label{d4.1}
\end{equation}%
where $\mathbf{x}=\mathbf{x}\left( t,\mathbf{\xi }\right) $, $\mathbf{u}=%
\mathbf{u}\left( t,\mathbf{x}\right) $. After elimination of the variable $%
\mathbf{u}$ we obtain instead of (\ref{d1.10})%
\begin{equation}
m\frac{d^{2}\mathbf{x}}{dt^{2}}=-\mathbf{\nabla }V\left( \mathbf{x}\right) -%
\mathbf{\nabla }U_{\mathrm{B}}  \label{d4.2}
\end{equation}%
where%
\begin{equation}
U_{\mathrm{B}}=U\left( \rho ,\mathbf{\nabla }\rho ,\mathbf{\nabla
}^{2}\rho \right) =\frac{\hbar ^{2}}{8m\rho }\left( \frac{\left(
\mathbf{\nabla }\rho
\right) ^{2}}{\rho }-2\mathbf{\nabla }^{2}\rho \right) =\mathbf{-}\frac{%
\hbar ^{2}}{2m\sqrt{\rho }}\mathbf{\nabla }^{2}\sqrt{\rho }
\label{d4.3}
\end{equation}%
\begin{equation}
\rho =\rho _{0}\left( \mathbf{\xi }\right) \left( \frac{\partial
\left( x^{1},x^{2},x^{3}\right) }{\partial \left( \xi _{1},\xi
_{2},\xi _{3}\right) }\right) ^{-1}  \label{d4.3a}
\end{equation}

To eliminate differentiation with respect to $\mathbf{x}$ and to
write
dynamic equations (\ref{d4.2}) in terms of the independent variables $t,%
\mathbf{\xi }$, we introduce the variable
\begin{equation}
R=\frac{\rho _{0}\left( \mathbf{\xi }\right) }{\rho
}=\frac{\partial \left( x^{1},x^{2},x^{3}\right) }{\partial \left(
\xi _{1},\xi _{2},\xi _{3}\right) }=\det \left\vert \left\vert
x^{\alpha ,\beta }\right\vert \right\vert ,\qquad \alpha ,\beta
=1,2,3  \label{d4.4}
\end{equation}%
as a multilinear function of variables $x^{\alpha ,\beta }\equiv
\partial x^{\alpha }/\partial \xi _{\beta }$. We take into account
that
\begin{equation}
\frac{\partial }{\partial x^{\alpha }}=\frac{\partial \xi _{\beta }}{%
\partial x^{\alpha }}\frac{\partial }{\partial \xi _{\beta }}=\frac{1}{R}%
\frac{\partial R}{\partial x^{\alpha ,\beta }}\frac{\partial
}{\partial \xi _{\beta }},  \label{d4.5}
\end{equation}%
Then we obtain dynamic equations (\ref{d4.2}) in the form%
\begin{equation}
m\ddot{x}^{\alpha }=-\frac{\partial V\left( \mathbf{x}\right)
}{\partial x^{\alpha }}+\frac{\hbar ^{2}}{2mR}\frac{\partial
R}{\partial x^{\alpha
,\beta }}\frac{\partial }{\partial \xi _{\beta }}\left[ \frac{1}{\sqrt{R}}%
\frac{\partial R}{\partial x^{\mu ,\nu }}\frac{\partial }{\partial
\xi _{\nu
}}\left( \frac{1}{R}\frac{\partial R}{\partial x^{\mu ,\sigma }}\frac{%
\partial }{\partial \xi _{\sigma }}\frac{1}{\sqrt{R}}\right) \right]  \label{d4.6}
\end{equation}%
In terms of independent variables $t,\mathbf{\xi }$ the mean value $\mathbf{u%
}$ of the stochastic velocity has the form
\begin{equation}
u^{\alpha }\left( t,\mathbf{\xi }\right) =-\frac{\hbar }{2m\rho
_{0}\left(
\mathbf{\xi }\right) }\frac{\partial R}{\partial x^{\alpha ,\beta }}\frac{%
\partial }{\partial \xi _{\beta }}\frac{\rho _{0}\left( \mathbf{\xi }\right)
}{R}  \label{d4.7}
\end{equation}

Thus, in the Lagrangian variables $t,\mathbf{\xi }$ the dynamic
equations for the statistical ensemble of stochastic (quantum)
particles are rather bulky. However, if the particles are
deterministic, and $\hbar =0$, dynamic equations (\ref{d4.6}) turn
to the ordinary differential equations
\begin{equation}
m\mathbf{\ddot{x}}=-\mathbf{\nabla }V\left( \mathbf{x}\right)
,\qquad \mathbf{x}=\mathbf{x}\left( t,\mathbf{\xi }\right)
\label{d4.8}
\end{equation}

If in the dynamic equations (\ref{d4.8}) the variable $\mathbf{x}$
does not depend on $\mathbf{\xi }$, they are dynamic equations for
the single classical particle. In order to pass from the equation
(\ref{d4.8}) for the single particle, described by
$\mathbf{x}=\mathbf{x}\left( t\right) $, to the dynamic equations
(\ref{d4.6}), i.e. "to quantize the classical
particle", one needs to consider statistical ensemble (replace $\mathbf{x}=%
\mathbf{x}\left( t\right) $ by $\mathbf{x}=\mathbf{x}\left( t,\mathbf{\xi }%
\right) $) and to add two last terms, containing the quantum
constant. Thus, the conventional quantization may be considered as
some dynamic procedure, introducing additional terms in the action
of the statistical ensemble. One needs no quantum principles for
such a quantization, because the concept of the wave function does
not used here, (the quantum principles are needed only for
explanation, what is the wave function)

If the relation (\ref{d1.10a}) takes place, and relations $\mathbf{x}=%
\mathbf{x}\left( t,\mathbf{\xi }\right) $ can be resolved with
respect to
variables $\mathbf{\xi }$ in the form $\mathbf{\xi }=x\left( t,\mathbf{x}%
\right) $, dynamic equations (\ref{d4.8}) can be rewritten in the
Eulerian
variables in the form (\ref{d1.14}), (\ref{d1.15})%
\begin{eqnarray}
\frac{\partial \mathbf{v}}{\partial t}+\left( \mathbf{v\nabla
}\right)
\mathbf{v} &=&-\frac{1}{m}\mathbf{\nabla }\left( U\left( \rho ,\mathbf{%
\nabla }\rho ,\mathbf{\nabla }^{2}\rho \right) +V\left(
\mathbf{x}\right)
\right)  \label{d4.9} \\
\frac{\partial \rho }{\partial t}+\mathbf{\nabla }\left( \rho \mathbf{v}%
\right) &=&0  \label{d4.10}
\end{eqnarray}%
where $\rho =\rho \left( t,\mathbf{x}\right) $,
$\mathbf{v}=\mathbf{v}\left( t,\mathbf{x}\right) $. In the
Eulerian coordinates the dynamic equations for the statistical
ensemble are simpler, than those in the Lagrangian coordinates. At
the same time they are rather demonstrable. To obtain the mean
trajectories of stochastic particles, one needs to solve at first
the dynamic equations (\ref{d4.9}), (\ref{d4.10}). When the
variables $\rho
=\rho \left( t,\mathbf{x}\right) $, $\mathbf{v}=\mathbf{v}\left( t,\mathbf{x}%
\right) $ become known, one needs to solve ordinary differential equations%
\begin{equation}
\frac{d\mathbf{x}}{dt}=\mathbf{v}\left( t,\mathbf{x}\right)
\label{d4.11}
\end{equation}

To obtain dynamic equations in terms of the generalized stream
function (the
Hamilton-Jacobi form), we are to integrate dynamic equations (\ref{d4.9}), (%
\ref{d4.10}) and formulate dynamic equations in terms of
hydrodynamic potentials $\mathbf{\xi }$ (Clebsch potentials
\cite{C57,C59}). The hydrodynamic potentials $\mathbf{\xi }$ may
be considered as the generalized stream function, because they
have the property of the stream function: (1) they label the world
lines of the fluid particles and (2) some combination of the
derivatives of $\mathbf{\xi }$ satisfy the continuity equation
identically at any values of $\mathbf{\xi }$. (See for details
\cite{R2004}).

To produce integration of dynamic equations, we return to the action (\ref%
{d4.1}), which has now the form%
\begin{equation}
\mathcal{A}_{\mathcal{E}\left[ \mathcal{S}_{\mathrm{st}}\right]
}\left[
\mathbf{x}\right] =\int \int\limits_{V_{\mathbf{\xi }}}\left\{ \frac{m}{2}%
\left( \frac{d\mathbf{x}}{dt}\right) ^{2}-V\left( \mathbf{x}\right) -U_{%
\mathrm{B}}\right\} \rho _{0}\left( \mathbf{\xi }\right)
dtd\mathbf{\xi } \label{d4.12}
\end{equation}%
where $\mathbf{x\equiv x}\left( t,\mathbf{\xi }\right) $. The variables $%
\rho $ and $U_{\mathrm{B}}=U\left( \rho ,\mathbf{\nabla }\rho ,\mathbf{%
\nabla }^{2}\rho \right) $ are defined by the relation (\ref{d4.3}), (\ref%
{d4.3a}).

To transform the action (\ref{d4.12}) to independent variables
$x=\left\{ x^{k}\right\} =\left\{ t,\mathbf{x}\right\} $, we use
the parametric
representation of the mean world lines $\mathbf{x\equiv x}\left( t,\mathbf{%
\xi }\right) $. Let
\begin{equation}
x^{k}=x^{k}\left( \xi _{0},\mathbf{\xi }\right) =x^{k}\left( \xi
\right) ,\qquad k=0,1,2,3  \label{d4.12a}
\end{equation}%
where $\xi =\left\{ \xi _{k}\right\} =\left\{ \xi _{0},\mathbf{\xi
}\right\} $, $k=0,1,2,3$. The shape of the world line is described
by $x^{k}$,
considered as a function of $\xi _{0}$ at fixed $\mathbf{\xi }$. The action (%
\ref{d4.12}) can be rewritten in the form
\begin{equation}
\mathcal{A}_{\mathcal{E}\left[ \mathcal{S}_{\mathrm{st}}\right] }\left[ x%
\right] =\int\limits_{V_{\mathbf{\xi }}}\left\{ \frac{m}{2}\left( \frac{%
\partial \mathbf{x}}{\partial \xi _{0}}\right) ^{2}\left( \frac{\partial
x^{0}}{\partial \xi _{0}}\right) ^{-1}-\left( V\left( x\right) +U_{\mathrm{B}%
}\right) \frac{\partial x^{0}}{\partial \xi _{0}}\right\} \rho
_{0}\left( \mathbf{\xi }\right) d^{4}\xi ,  \label{d4.14}
\end{equation}

Let us consider the variables $\xi =\left\{ \xi _{k}\right\} $,
$k=0,1,2,3$ as dependent variables and variables $x=\left\{
x^{k}\right\} $ as independent ones. We consider the Jacobian
\begin{equation}
J=\frac{\partial \left( \xi _{0}\mathbf{,}\xi _{1},\xi _{2},\xi
_{3}\right) }{\partial \left( x^{0},x^{1},x^{2},x^{3}\right)
}=\det \left\vert
\left\vert \xi _{l,k}\right\vert \right\vert ,\qquad \xi _{l,k}\equiv \frac{%
\partial \xi _{l}}{\partial x^{k}}\qquad l,k=0,1,2,3  \label{d4.15}
\end{equation}%
as a four-linear function of variables $\xi _{l,k}\equiv \partial
_{k}\xi _{l}$, $l,k=0,1,2,3$. We take into account that
\begin{equation}
\frac{\partial x^{k}}{\partial \xi _{0}}=\frac{\partial \left( x^{k}\mathbf{,%
}\xi _{1},\xi _{2},\xi _{3}\right) }{\partial \left( \xi
_{0}\mathbf{,}\xi _{1},\xi _{2},\xi _{3}\right) }=\frac{\partial
\left( x^{k}\mathbf{,}\xi _{1},\xi _{2},\xi _{3}\right) }{\partial
\left( x^{0},x^{1},x^{2},x^{3}\right) }\frac{\partial \left(
x^{0},x^{1},x^{2},x^{3}\right) }{\partial \left( \xi
_{0}\mathbf{,}\xi _{1},\xi _{2},\xi _{3}\right)
}=J^{-1}\frac{\partial J}{\partial \xi _{0,k}} \label{d4.16}
\end{equation}%
and%
\begin{equation}
\rho =\rho _{0}\left( \mathbf{\xi }\right) \frac{\partial \left(
\xi
_{1},\xi _{2},\xi _{3}\right) }{\partial \left( x^{1},x^{2},x^{3}\right) }%
=\rho _{0}\left( \mathbf{\xi }\right) \frac{\partial \left( x^{0}\mathbf{,}%
\xi _{1},\xi _{2},\xi _{3}\right) }{\partial \left(
x^{0},x^{1},x^{2},x^{3}\right) }=\rho _{0}\left( \mathbf{\xi }\right) \frac{%
\partial J}{\partial \xi _{0,0}}  \label{d4.17}
\end{equation}

The action (\ref{d4.14}) takes the form%
\begin{equation}
\mathcal{A}_{\mathcal{E}\left[ \mathcal{S}_{\mathrm{st}}\right] }\left[ \xi %
\right] =\int\limits_{V_{x}}\left\{ \frac{m}{2}\left( \frac{\partial J}{%
\partial \xi _{0,\alpha }}\right) ^{2}\left( \frac{\partial J}{\partial \xi
_{0,0}}\right) ^{-2}-V\left( x\right) -U_{\mathrm{B}}\right\} \rho
d^{4}x \label{d4.18}
\end{equation}%
\[
\rho \equiv \rho _{0}\left( \mathbf{\xi }\right) \frac{\partial
J}{\partial \xi _{0,0}}
\]%
It follows from (\ref{d1.10}) that%
\begin{equation}
\rho U_{\mathrm{B}}=\rho U\left( \rho ,\mathbf{\nabla }\rho ,\mathbf{\nabla }%
^{2}\rho \right) =\frac{\hbar ^{2}}{8m}\frac{\left( \mathbf{\nabla
}\rho \right) ^{2}}{\rho }-\frac{\hbar ^{2}}{4m}\mathbf{\nabla
}^{2}\rho \label{d4.18a}
\end{equation}%
The last term of (\ref{d4.18a}) has a form of divergence, and it
does not contribute to dynamic equations. This term may be
omitted.

If the relation (\ref{d1.10a})
\begin{equation}
\frac{\partial J}{\partial \xi _{0,0}}\neq 0  \label{d4.18b}
\end{equation}%
takes place, the variational problems (\ref{d4.14}) and
(\ref{d4.18}) are equivalent. On the contrary, if the relation
(\ref{d4.18b}) is violated we cannot be sure, that they are
equivalent.

Now we introduce designation $j=\left\{ j^{0},\mathbf{j}\right\}
=\left\{ \rho ,\mathbf{j}\right\} =\left\{ j^{k}\right\} $,
$k=0,1,2,3$
\begin{equation}
j^{k}=\rho _{0}\left( \mathbf{\xi }\right) \frac{\partial \left( x^{k}%
\mathbf{,}\xi _{1},\xi _{2},\xi _{3}\right) }{\partial \left(
x^{0},x^{1},x^{2},x^{3}\right) }=\rho _{0}\left( \mathbf{\xi }\right) \frac{%
\partial J}{\partial \xi _{0,k}},\qquad k=0,1,2,3  \label{d4.19}
\end{equation}%
and add designation (\ref{d4.19}) to the action (\ref{d4.18}) by
means the
Lagrangian multipliers $p_{k}$, $k=0,1,2,3$. We obtain%
\begin{equation}
\mathcal{A}_{\mathcal{E}\left[ \mathcal{S}_{\mathrm{st}}\right]
}\left[ \xi
,j,p\right] =\int\limits_{V_{x}}\left\{ m\frac{\mathbf{j}^{2}}{2\rho }%
-V\left( x\right) \rho -\rho U_{\mathrm{B}}-p_{k}\left( j^{k}-\rho
_{0}\left( \mathbf{\xi }\right) \frac{\partial J}{\partial \xi _{0,k}}%
\right) \right\} d^{4}x,  \label{d4.20}
\end{equation}%
\[
U_{\mathrm{B}}=U\left( \rho ,\mathbf{\nabla }\rho ,\mathbf{\nabla
}^{2}\rho \right) ,\qquad \rho \equiv j^{0}
\]%
Note that the action (\ref{d4.18}) and the action (\ref{d4.20})
describe the same variational problem. The action (\ref{d4.20}) is
interesting in the
sense, that the Lagrangian coordinates $\xi =\left\{ \xi _{0},\mathbf{\xi }%
\right\} $ are concentrated in the last term of the action. The
Lagrangian coordinates $\xi =\left\{ \xi _{0},\mathbf{\xi
}\right\} $ are defined to
within the transformation%
\begin{equation}
\xi _{0}=f_{0}\left( \tilde{\xi}_{0}\right) \mathbf{,\qquad }\xi
_{\alpha }=f_{\alpha }\left( \mathbf{\tilde{\xi}}\right) ,\qquad
\alpha =1,2,3 \label{d4.21}
\end{equation}%
where $f_{k}$, $k=0,1,2,3$ are arbitrary functions. The variable
$\xi _{0}$ is fictitious, and variation with respect to $\xi _{0}$
does not give an independent dynamic equation.

Variation of the action (\ref{d4.20}) with respect to $\xi _{l}$,
$l=0,1,2,3$ leads to the dynamic equations
\begin{equation}
\frac{\delta \mathcal{A}_{\mathcal{E}\left[
\mathcal{S}_{\mathrm{st}}\right]
}}{\delta \xi _{l}}=-\partial _{s}\left( \rho _{0}\left( \mathbf{\xi }%
\right) p_{k}\frac{\partial ^{2}J}{\partial \xi _{0,k}\partial \xi _{l,s}}%
\right) +p_{k}\frac{\partial \rho _{0}}{\partial \xi _{l}}\left(
\mathbf{\xi }\right) \frac{\partial J}{\partial \xi
_{0,k}}=0,\qquad l=0,1,2,3 \label{d4.22}
\end{equation}%
Using identities%
\begin{equation}
\frac{\partial J}{\partial \xi _{i,l}}\xi _{k,l}\equiv J\delta
_{k}^{i},\qquad \partial _{l}\frac{\partial ^{2}J}{\partial \xi
_{0,k}\partial \xi _{i,l}}\equiv 0  \label{A.10}
\end{equation}%
\begin{equation}
\frac{\partial ^{2}J}{\partial \xi _{0,k}\partial \xi
_{l,s}}\equiv
J^{-1}\left( \frac{\partial J}{\partial \xi _{0,k}}\frac{\partial J}{%
\partial \xi _{l,s}}-\frac{\partial J}{\partial \xi _{0,s}}\frac{\partial J}{%
\partial \xi _{l,k}}\right)  \label{A.9}
\end{equation}%
we obtain from (\ref{d4.22})%
\[
-\frac{\partial ^{2}J}{\partial \xi _{0,k}\partial \xi _{l,s}}\rho
_{0}\partial _{s}p_{k}-\frac{\partial ^{2}J}{\partial \xi
_{0,k}\partial \xi
_{l,s}}\frac{\partial \rho _{0}}{\partial \xi _{j}}\xi _{j,s}p_{k}+p_{k}%
\frac{\partial \rho _{0}}{\partial \xi _{l}}\frac{\partial
J}{\partial \xi _{0,k}}=0,\qquad l=0,1,2,3
\]%
\begin{eqnarray}
&&-J^{-1}\left( \frac{\partial J}{\partial \xi _{0,k}}\frac{\partial J}{%
\partial \xi _{l,s}}-\frac{\partial J}{\partial \xi _{0,s}}\frac{\partial J}{%
\partial \xi _{l,k}}\right) \rho _{0}\left( \mathbf{\xi }\right) \partial
_{s}p_{k}-\left( \frac{\partial J}{\partial \xi _{0,k}}\delta
_{j}^{l}-\delta _{j}^{0}\frac{\partial J}{\partial \xi _{l,k}}\right) \frac{%
\partial \rho _{0}\left( \mathbf{\xi }\right) }{\partial \xi _{j}}p_{k}
 \nonumber\\
&&\qquad\qquad+p_{k}\frac{\partial \rho _{0}\left( \mathbf{\xi
}\right) }{\partial \xi _{l}}\frac{\partial J}{\partial \xi
_{0,k}}=0,\qquad l=0,1,2,3\label{d4.22a}
\end{eqnarray}%
Simplifying (\ref{d4.22a}) by means of the first identity
(\ref{A.10}), we
obtain%
\begin{equation}
J^{-1}\left( \frac{\partial J}{\partial \xi _{0,k}}\frac{\partial J}{%
\partial \xi _{l,s}}-\frac{\partial J}{\partial \xi _{0,s}}\frac{\partial J}{%
\partial \xi _{l,k}}\right) \rho _{0}\partial _{s}p_{k}=0  \label{d4.22b}
\end{equation}

Convoluting (\ref{d4.22b}) with $\xi _{l,i}$ and using the first identity (%
\ref{A.10}) and designations (\ref{d4.19}), we obtain
\begin{equation}
j^{k}\partial _{i}p_{k}-j^{k}\partial _{k}p_{i}=0,\qquad i=0,1,2,3
\label{A.10a}
\end{equation}

Variation of (\ref{d4.20}) with respect to $j^{\beta }$ gives
\begin{equation}
p_{\beta }=m\frac{j^{\beta }}{\rho },\qquad \beta =1,2,3
\label{A.17}
\end{equation}%
Variating (\ref{d4.20}) with respect to $j^{0}=\rho $, using designations%
\[
\rho _{\gamma }\equiv \partial _{\gamma }\rho ,\qquad \rho
_{\alpha \beta }\equiv \partial _{\alpha }\partial _{\beta }\rho
\]%
and taking into account relation (\ref{d4.3}) for
$U_{\mathrm{B}}=U\left(
\rho ,\mathbf{\nabla }\rho ,\mathbf{\nabla }^{2}\rho \right) $, we obtain%
\begin{eqnarray}
p_{0} &=&-\frac{m}{2\rho ^{2}}j^{\alpha }j^{\alpha }-V\left( x\right) -\frac{%
\partial }{\partial \rho }\left( \rho U_{\mathrm{B}}\right) +\mathbf{%
\partial }_{\gamma }\frac{\partial }{\partial \rho _{\gamma }}\left( \rho U_{%
\mathrm{B}}\right) -\partial _{\alpha }\partial _{\beta }\frac{\partial }{%
\partial \rho _{\alpha \beta }}\left( \rho U_{\mathrm{B}}\right)  \nonumber
\\
&=&-\frac{m}{2\rho ^{2}}j^{\alpha }j^{\alpha }-V\left( x\right) -U_{\mathrm{B%
}}  \label{A.10b}
\end{eqnarray}

We note the remarkable property of the Bohm potential $U_{\mathrm{B}%
}=U\left( \rho ,\mathbf{\nabla }\rho ,\mathbf{\nabla }^{2}\rho
\right) $,
defined by the relation (\ref{d4.3}). The quantity $p_{0}$ is expressed via $%
U_{\mathrm{B}}=U\left( \rho ,\mathbf{\nabla }\rho ,\mathbf{\nabla
}^{2}\rho
\right) $ in such a way, as if $U\left( \rho ,\mathbf{\nabla }\rho ,\mathbf{%
\nabla }^{2}\rho \right) $ does not depend on $\rho $ and its
derivatives.

Eliminating $p_{k}$ from the equations (\ref{A.10a}) by means of relations (%
\ref{A.17}), (\ref{A.10b}) and setting $\mathbf{v}=\mathbf{j}/\rho
$, we obtain dynamic equations in the Eulerian form (\ref{d4.9}).

There is another possibility. The dynamic equations (\ref{d4.22b})
may be considered to be linear partial differential equations with
respect to variables $p_{k}$. They can be solved in the form
\begin{equation}
p_{k}=b_{0}\left( \partial _{k}\varphi +g^{\alpha }\left( \mathbf{\xi }%
\right) \partial _{k}\xi _{\alpha }\right) ,\qquad k=0,1,2,3
\label{g1.30}
\end{equation}%
where $g^{\alpha }\left( \mathbf{\xi }\right) ,\;\;\alpha =1,2,3$
are arbitrary functions of the argument $\mathbf{\xi }=\left\{ \xi
_{1},\xi
_{2},\xi _{3}\right\} $, $b_{0}\neq 0$ is an arbitrary real constant, and $%
\varphi $ is the variable $\xi _{0}$, which ceases to be
fictitious. Note
that the constant $b_{0}$ may be eliminated, including it in the functions $%
\mathbf{g}=\left\{ g^{1},g^{2},g^{3}\right\} $ and in the variable $\varphi $%
.

One can test by the direct substitution that the relation
(\ref{g1.30}) is
the general solution of linear equations (\ref{d4.22b}). Substituting (\ref%
{g1.30}) in (\ref{d4.22b}) and taking into account antisymmetry of
the
bracket in (\ref{d4.22b}) with respect to transposition of indices $k$ and $%
s $, we obtain
\begin{equation}
J^{-1}\rho _{0}\left( \mathbf{\xi }\right) \left( \frac{\partial
J}{\partial \xi _{0,k}}\frac{\partial J}{\partial \xi
_{l,s}}-\frac{\partial J}{\partial \xi _{0,s}}\frac{\partial
J}{\partial \xi _{l,k}}\right) \frac{\partial g^{\alpha }\left(
\mathbf{\xi }\right) }{\partial \xi _{\mu }}\xi _{\mu ,s}\xi
_{\alpha ,k}=0  \label{g1.32}
\end{equation}%
The relation (\ref{g1.32}) is the valid equality, as it follows
from the first identity (\ref{A.10}).

Let us substitute (\ref{g1.30}) in the action (\ref{d4.20}).
Taking into account the first identity (\ref{A.10}) and omitting
the term
\[
\rho _{0}\left( \mathbf{\xi }\right) \frac{\partial J}{\partial \xi _{0,k}}%
\partial _{k}\varphi =\rho _{0}\left( \mathbf{\xi }\right) \frac{\partial
\left( \varphi ,\xi _{1},\xi _{2},\xi _{3}\right) }{\partial
\left( x^{0},x^{1},x^{2},x^{3}\right) }
\]%
which does not contribute to the dynamic equations, we obtain
\begin{equation}
\mathcal{A}_{\mathcal{E}\left[ \mathcal{S}_{\mathrm{st}}\right]
}\left[ \varphi ,\mathbf{\xi },j\right] =\int \left\{
\frac{m}{2}\frac{j^{\alpha }j^{\alpha }}{j^{0}}-V\left( x\right)
\rho -U_{\mathrm{B}}\rho
-j^{k}b_{0}\left( \partial _{k}\varphi +g^{\alpha }\left( \mathbf{\xi }%
\right) \partial _{k}\xi _{\alpha }\right) \right\}
d^{4}x\mathbf{,} \label{d4.25}
\end{equation}%
\[
j^{0}\equiv \rho
\]

Variation of (\ref{d4.25}) with respect to $j^{0}\equiv \rho $ gives%
\begin{equation}
-\frac{m\mathbf{j}^{2}}{2\rho ^{2}}-V\left( x\right) -U_{\mathrm{B}%
}-b_{0}\left( \partial _{0}\varphi +g^{\alpha }\left( \mathbf{\xi
}\right)
\partial _{0}\xi _{\alpha }\right) =0,\qquad U_{\mathrm{B}}=\frac{\hbar ^{2}%
}{8m}\left( \frac{\left( \mathbf{\nabla }\rho \right) ^{2}}{\rho ^{2}}-2%
\frac{\mathbf{\nabla }^{2}\rho }{\rho }\right)  \label{d4.26}
\end{equation}%
Variation with respect to $j^{\mu }$ gives%
\begin{equation}
m\frac{j^{\mu }}{\rho }=b_{0}\left( \partial _{\mu }\varphi
+g^{\alpha }\left( \mathbf{\xi }\right) \partial _{\mu }\xi
_{\alpha }\right) \label{d4.26a}
\end{equation}%
Variation with respect to $\varphi $ gives%
\begin{equation}
\partial _{k}j^{k}=0  \label{d4.27}
\end{equation}

Finally, varying (\ref{d4.25}) with respect to $\xi _{\mu }$ and
taking into account (\ref{d4.27}), we obtain
\begin{equation}
b_{0}j^{k}\Omega ^{\alpha \mu }\left( \mathbf{\xi }\right)
\partial _{k}\xi
_{\alpha }=0,\qquad \Omega ^{a\mu }\left( \mathbf{\xi }\right) =\left( \frac{%
\partial g^{\alpha }\left( \mathbf{\xi }\right) }{\partial \xi _{\mu }}-%
\frac{\partial g^{\mu }\left( \mathbf{\xi }\right) }{\partial \xi _{\alpha }}%
\right)  \label{d4.28}
\end{equation}

If
\begin{equation}
\det \left\vert \left\vert \Omega ^{\alpha \mu }\right\vert
\right\vert \neq 0  \label{d4.28c}
\end{equation}%
then taking into account that the velocity
$\mathbf{v}=\mathbf{j}/j^{0}$,
one obtains from (\ref{d4.28}), so called Lin constraint \cite{L63}%
\begin{equation}
\partial _{0}\mathbf{\xi +}\left( \mathbf{v\nabla }\right) \mathbf{\xi }=0
\label{d4.28b}
\end{equation}%
which means that the variables $\mathbf{\xi }=\left\{ \xi _{1},\xi
_{2},\xi _{3}\right\} $ are constant along the mean world lines of
particles. In other words, the variables $\mathbf{\xi }$ are
Lagrangian coordinates, which label mean world lines of particles.

However, the constraint (\ref{d4.28c}) is not takes place always.
In particular, $\Omega ^{\alpha \beta }\equiv 0$ in the case of
irrotational flow. Besides, the quantity $\Omega ^{\alpha \beta }$
is antisymmetric, as it follows from the second relation
(\ref{d4.28}), and
\begin{equation}
\det \left\vert \left\vert \Omega ^{\alpha \beta }\right\vert
\right\vert =\left\vert
\begin{array}{ccc}
0 & \Omega ^{12} & \Omega ^{13} \\
-\Omega ^{12} & 0 & \Omega ^{23} \\
-\Omega ^{13} & -\Omega ^{23} & 0%
\end{array}%
\right\vert \equiv 0  \label{d4.28a}
\end{equation}%
Note that identity (\ref{d4.28a}) is a property of the
three-dimensional space. In the two-dimensional space $\det
\left\vert \left\vert \Omega ^{\alpha \beta }\right\vert
\right\vert =\left( \Omega ^{12}\right) ^{2}$. In the case of
four-dimensional space we have
\[
\det \left\vert \left\vert \Omega ^{\alpha \beta }\right\vert
\right\vert =\left( \Omega ^{12}\Omega ^{34}-\Omega ^{13}\Omega
^{24}+\Omega ^{14}\Omega ^{23}\right) ^{2}
\]

It seems rather strange and unexpected, that the Lin constraint (\ref{d4.28b}%
) is not a corollary of the dynamic equation (\ref{d4.28}),
although the Lin
constraint (\ref{d4.28b}) it is compatible with the dynamic equation (\ref%
{d4.28}). In the case of nonrotational flow the Euler hydrodynamic
equations for perfect fluid can be obtained from the variational
principle \cite{D49}. In the case of rotational flow of the same
fluid the Euler hydrodynamic equations can be deduced from the
variational principle, only when the Lin constraints are
introduced in the action functional as side conditions, and
the variables $\mathbf{\xi }$ are considered as dynamic variables \cite{L63}%
. Does it mean, that the Lagrangian coordinates $\mathbf{\xi }$
are inadequate dynamical variables? Maybe. It is not clear now.

From equations (\ref{d4.26}) - (\ref{d4.28b}) one obtains five equations%
\begin{equation}
-\frac{\left( \mathbf{\nabla }\varphi +g^{\alpha }\left( \mathbf{\xi }%
\right) \mathbf{\nabla }\xi _{\alpha }\right) ^{2}}{2m}-V\left( x\right) -U_{%
\mathrm{B}}-\left( \partial _{0}\varphi +g^{\alpha }\left( \mathbf{\xi }%
\right) \partial _{0}\xi _{\alpha }\right) =0,\qquad
\label{b4.29}
\end{equation}%
\begin{equation}
\partial _{0}\mathbf{\xi +}\left( \mathbf{v\nabla }\right) \mathbf{\xi }=0
\label{b4.30}
\end{equation}%
\begin{equation}
\partial _{0}\rho +\mathbf{\nabla }\left( \rho \frac{\left( \mathbf{\nabla }%
\varphi +g^{\alpha }\left( \mathbf{\xi }\right) \mathbf{\nabla
}\xi _{\alpha }\right) }{m}\right)  \label{b4.31}
\end{equation}%
for five dynamic variables $\rho ,\varphi ,\mathbf{\xi
}$\textbf{.} Indefinite functions $\mathbf{g}\left( \mathbf{\xi
}\right) =\{g^{1}\left( \mathbf{\xi }\right) $,$g^{2}\left(
\mathbf{\xi }\right) ,g^{3}\left( \mathbf{\xi }\right) \}$ are
determined from initial conditions for velocity
$\mathbf{v}=\mathbf{j}/\rho $. The constant $b_{0}$ is included in
the indefinite functions $\varphi ,\mathbf{g}\left( \mathbf{\xi
}\right) $ The
velocity $\mathbf{v}$ is expressed via dynamic variables $\rho ,\varphi ,%
\mathbf{\xi }$ by means of the relation
\begin{equation}
\mathbf{v}=\frac{\mathbf{j}}{\rho }=\frac{\left( \mathbf{\nabla
}\varphi +g^{\alpha }\left( \mathbf{\xi }\right) \mathbf{\nabla
}\xi _{\alpha }\right) }{m}  \label{b4.32}
\end{equation}

\section{Meaning of functions $\mathbf{g=}\left\{ g^{1},g^{2},g^{3}\right\} $%
}

The arbitrary functions $\mathbf{g=}\left\{ g^{1}\left(
\mathbf{\xi }\right) ,g^{2}\left( \mathbf{\xi }\right)
,g^{3}\left( \mathbf{\xi }\right) \right\}
$ may be derived from initial values of hydrodynamic variables $\rho ,$ $%
\mathbf{v}$. Let at the initial moment $t=0$%
\begin{equation}
\rho \left( 0,\mathbf{x}\right) =\rho _{\mathrm{in}}\left(
\mathbf{x}\right) ,\qquad \mathbf{v}\left( 0,\mathbf{x}\right)
=\mathbf{v}_{\mathrm{in}}\left( \mathbf{x}\right)  \label{d5.1}
\end{equation}%
Let us choose the initial form of labelling in the form%
\begin{equation}
\mathbf{\xi }\left( 0,\mathbf{x}\right) =\mathbf{\xi
}_{\mathrm{in}}\left( \mathbf{x}\right) =\mathbf{x,\qquad }\varphi
\left( 0,\mathbf{x}\right) =\varphi _{\mathrm{in}}\left(
\mathbf{x}\right) =0  \label{d5.2}
\end{equation}%
Setting $t=0$ in (\ref{b4.32}), (\ref{d4.17}) and taking into account (\ref%
{d5.1}) and (\ref{d5.2}), we obtain respectively

\begin{equation}
\mathbf{v}\left( 0,\mathbf{x}\right) =\mathbf{v}_{\mathrm{in}}\left( \mathbf{%
x}\right) =\frac{\mathbf{g}\left( \mathbf{x}\right) }{m}
\label{d5.3}
\end{equation}%
\begin{equation}
\rho \left( 0,\mathbf{x}\right) =\rho _{\mathrm{in}}\left(
\mathbf{x}\right)
=\rho _{0}\left( \mathbf{x}\right) \frac{\partial \left( \xi _{\mathrm{in}%
1}\left( \mathbf{x}\right) ,\xi _{\mathrm{in}2}\left(
\mathbf{x}\right) ,\xi _{\mathrm{in}3}\left( \mathbf{x}\right)
\right) }{\partial \left( x^{1},x^{2},x^{3}\right) }=\rho
_{0}\left( \mathbf{x}\right)  \label{d5.4}
\end{equation}%
Thus, arbitrary functions $\mathbf{g}\left( \mathbf{\xi }\right) $
and the weight function $\rho _{0}\left( \mathbf{\xi }\right) $
may be uniquely
determined via initial values $\rho _{\mathrm{in}}\left( \mathbf{x}\right) $%
, $\mathbf{v}_{\mathrm{in}}\left( \mathbf{x}\right) $ of
quantities $\rho $, $\mathbf{v}$.

Eliminating functions $\mathbf{g}\left( \mathbf{\xi }\right) $
from dynamic equations (\ref{b4.29}) - (\ref{b4.31}) by means of
relations (\ref{d5.3}),
we obtain%
\begin{equation}
\partial _{0}\xi _{\alpha }+\left( \frac{1}{m}\mathbf{\nabla }\varphi +%
\mathbf{v}_{\mathrm{in}}\left( \mathbf{\xi }\right) \right) \mathbf{\nabla }%
\xi _{\alpha }=0,\qquad \alpha =1,2,3  \label{d5.6}
\end{equation}%
\begin{equation}
\partial _{0}\varphi +\frac{\left( \mathbf{\nabla }\varphi \right) ^{2}}{2m}+%
\frac{m\left( \mathbf{v}_{\mathrm{in}}\left( \mathbf{\xi }\right)
\partial
_{\alpha }\mathbf{\xi }\right) \left( \mathbf{v}_{\mathrm{in}}\left( \mathbf{%
\xi }\right) \partial _{\alpha }\mathbf{\xi }\right) }{2}-mv_{\mathrm{in}%
}^{\alpha }\left( \mathbf{\xi }\right)
\mathbf{v}_{\mathrm{in}}\left(
\mathbf{\xi }\right) \mathbf{\nabla }\xi _{\alpha }+V\left( x\right) +U_{%
\mathrm{B}}\left( \rho \right) =0  \label{d5.9}
\end{equation}%
\begin{equation}
\partial _{0}\rho +\mathbf{\nabla }\left( \rho \left( \frac{1}{m}\mathbf{%
\nabla }\varphi +v_{\mathrm{in}}^{\beta }\left( \mathbf{\xi }\right) \mathbf{%
\nabla }\xi _{\beta }\right) \right) =0  \label{d5.8a}
\end{equation}

The initial values $\mathbf{\xi }_{\mathrm{in}}\left( \mathbf{x}\right) $%
\textbf{, }$\varphi _{\mathrm{in}}\left( \mathbf{x}\right) $ of
hydrodynamic potentials $\mathbf{\xi }$, $\varphi $ may be chosen
universally for all
flows, for instance, in the form (\ref{d5.2}). It means, that equations (\ref%
{b4.30}), (\ref{b4.31}) are essentially equations, describing the
labelling evolution at fixed dynamics.

\section{Description in terms of complex potential}

One may form complex potential $\psi $ from the Clebsch potentials $\mathbf{%
\xi }$, $\varphi $ and the density $\rho $. This complex potential
$\psi $ is known as the wave function, or $\psi $-function. By
means of a change of variables the action (\ref{d4.25}) can be
transformed to a description in terms of a wave function
\cite{R99}. Let us introduce the $k$-component complex function
$\psi =\{\psi _{\alpha }\},\;\;\alpha =1,2,...k$, defining it by
the relations
\begin{equation}
\psi _{\alpha }=\sqrt{\rho }e^{i\varphi }u_{\alpha }(\mathbf{\xi
}),\qquad
\psi _{\alpha }^{\ast }=\sqrt{\rho }e^{-i\varphi }u_{\alpha }^{\ast }(%
\mathbf{\xi }),\qquad \alpha =1,2,...k  \label{d6.1}
\end{equation}%
\begin{equation}
\psi ^{\ast }\psi \equiv \sum_{\alpha =1}^{k}\psi _{\alpha }^{\ast
}\psi _{\alpha }  \label{d6.2}
\end{equation}%
where (*) means the complex conjugate, $u_{\alpha }(\mathbf{\xi })$, $%
\;\alpha =1,2,...k$ are functions of only variables $\mathbf{\xi
}$. They satisfy the relations
\begin{equation}
-\frac{i}{2}\sum_{\alpha =1}^{k}(u_{\alpha }^{\ast }\frac{\partial
u_{\alpha }}{\partial \xi _{\beta }}-\frac{\partial u_{\alpha
}^{\ast }}{\partial \xi _{\beta }}u_{\alpha })=g^{\beta
}(\mathbf{\xi }),\qquad \beta =1,2,...k\qquad \sum_{\alpha
=1}^{k}u_{\alpha }^{\ast }u_{\alpha }=1 \label{d6.3}
\end{equation}%
where $k$ is such a natural number that equations (\ref{d6.3})
admit a
solution. In general, $k$ depends on the form of the arbitrary functions $%
\mathbf{g}=\{g^{\beta }(\mathbf{\xi })\}$,\ $\beta =1,2,3.$

It is easy to verify, that
\begin{equation}
\rho =\psi ^{\ast }\psi ,\qquad j^{\mu }=-\frac{ib_{0}}{2m}(\psi
^{\ast }\partial _{\mu }\psi -\partial _{\mu }\psi ^{\ast }\cdot
\psi ),\qquad \mu =1,2,3  \label{d6.4}
\end{equation}%
The variational problem with the action (\ref{d4.25}) appears to
be equivalent \cite{R99} to the variational problem with the
action functional
\begin{eqnarray}
\mathcal{A}[\psi ,\psi ^{\ast }] &=&\int \left\{
\frac{ib_{0}}{2}(\psi
^{\ast }\partial _{0}\psi -\partial _{0}\psi ^{\ast }\psi )+\frac{%
b_{0}^{2}(\psi ^{\ast }\mathbf{\nabla }\psi -\mathbf{\nabla }\psi
^{\ast
}\cdot \psi )^{2}}{8m\psi ^{\ast }\psi }\right.  \nonumber \\
&&\left. -\frac{\hbar ^{2}}{8m}\frac{\left( \mathbf{\nabla }\left(
\psi ^{\ast }\psi \right) \right) ^{2}}{\psi ^{\ast }\psi
}-V\left( x\right) \psi ^{\ast }\psi \right\} \mathrm{d}^{4}x
\label{d6.5}
\end{eqnarray}%
where $\mathbf{\nabla }=\left\{ \partial _{\alpha }\right\}
,\;\;\alpha =1,2,3$.

Let us consider the case, when the number $k$ of the wave function
components is equal to $2$. In this case the wave function $\psi
=\left\{ _{\psi _{2}}^{\psi _{1}}\right\} $ has four real
components. The number of hydrodynamic variables $\rho $,
$\mathbf{j}$ is also four, and we may hope that the first three
equations (\ref{d6.3}) can be solved for any choice of functions
$\mathbf{g}$. For the two-component wave function $\psi $ we have
the identity%
\begin{equation}
(\psi ^{\ast }\mathbf{\nabla }\psi -\mathbf{\nabla }\psi ^{\ast
}\cdot \psi )^{2}\equiv -4\rho \mathbf{\nabla }\psi ^{\ast }\cdot
\mathbf{\nabla }\psi +\left( \mathbf{\nabla }\rho \right)
^{2}+4\rho ^{2}\sum\limits_{\alpha =1}^{3}\left( \mathbf{\nabla
}s_{\alpha }\right) ^{2}  \label{d6.6}
\end{equation}%
where%
\begin{equation}
\rho =\psi ^{\ast }\psi ,\qquad s_{\alpha }=\frac{\psi ^{\ast
}\sigma _{\alpha }\psi }{\rho },\qquad \alpha =1,2,3  \label{d6.7}
\end{equation}%
$\sigma _{\alpha }$ are $2\times 2$ Pauli matrices%
\begin{equation}
\sigma _{1}=\left(
\begin{array}{cc}
0 & 1 \\
1 & 0%
\end{array}%
\right) ,\qquad \sigma _{2}=\left(
\begin{array}{cc}
0 & -i \\
i & 0%
\end{array}%
\right) ,\qquad \sigma _{3}=\left(
\begin{array}{cc}
1 & 0 \\
0 & -1%
\end{array}%
\right) ,  \label{d6.8}
\end{equation}%
Substituting (\ref{d6.6}) in (\ref{d6.5}), we obtain%
\begin{eqnarray}
\mathcal{A}[\psi ,\psi ^{\ast }] &=&\int \left\{
\frac{ib_{0}}{2}(\psi
^{\ast }\partial _{0}\psi -\partial _{0}\psi ^{\ast }\cdot \psi )-\frac{%
b_{0}^{2}}{2m}\mathbf{\nabla }\psi ^{\ast }\cdot \mathbf{\nabla }\psi +\frac{%
b_{0}^{2}}{8m}\rho \left( \mathbf{\nabla }s_{\alpha }\right)
^{2}\right.
\nonumber \\
&&\left. +\frac{b_{0}^{2}}{8m}\frac{\left( \mathbf{\nabla }\rho \right) ^{2}%
}{\rho }-\frac{\hbar ^{2}}{8m}\frac{\left( \mathbf{\nabla }\rho \right) ^{2}%
}{\rho }-V\left( x\right) \psi ^{\ast }\psi \right\}
\mathrm{d}^{4}x \label{d6.9}
\end{eqnarray}

If we choose the arbitrary constant $b_{0}$ in the form
$b_{0}=\hbar $, the
action (\ref{d6.9}) takes the form%
\begin{eqnarray}
\mathcal{A}[\psi ,\psi ^{\ast }] &=&\int \left\{ \frac{i\hbar
}{2}(\psi
^{\ast }\partial _{0}\psi -\partial _{0}\psi ^{\ast }\cdot \psi )-\frac{%
\hbar ^{2}}{2m}\mathbf{\nabla }\psi ^{\ast }\cdot \mathbf{\nabla
}\psi
\right.  \nonumber \\
&&+\left. \frac{\hbar ^{2}}{8m}\rho \mathbf{\nabla }s_{\alpha }\mathbf{%
\nabla }s_{\alpha }-V\left( x\right) \rho \right\} \mathrm{d}^{4}x
\label{d6.10}
\end{eqnarray}

In the case, when the wave function $\psi $ is one-component, for instance $%
\psi =\left\{ _{0}^{\psi _{1}}\right\} $, the quantities
$\mathbf{s=}\left\{
s_{1},s_{2},s_{3}\right\} $ are constant ($s_{1}=0,\ \ s_{2}=0,\ \ s_{3}=1$%
), the action (\ref{d6.10}) turns into
\begin{equation}
\mathcal{A}[\psi ,\psi ^{\ast }]=\int \left\{ \frac{i\hbar
}{2}(\psi ^{\ast
}\partial _{0}\psi -\partial _{0}\psi ^{\ast }\cdot \psi )-\frac{\hbar ^{2}}{%
2m}\mathbf{\nabla }\psi ^{\ast }\cdot \mathbf{\nabla }\psi
-V\left( x\right) \psi ^{\ast }\psi \right\} \mathrm{d}^{4}x
\label{d6.11}
\end{equation}%
The dynamic equation, generated by the action (\ref{d6.11}), is the Schr\"{o}%
dinger equation%
\begin{equation}
i\hbar \partial _{0}\psi +\frac{\hbar ^{2}}{2m}\mathbf{\nabla
}^{2}\psi -V\left( x\right) \psi =0  \label{d6.12}
\end{equation}%
This dynamic equation describes the flow of the fluid.

In the general case the dynamic equation, generated by the action
(\ref{d6.9}), has the form%
\begin{eqnarray}
&&ib_{0}\partial _{0}\psi +\frac{b_{0}^{2}}{2m}\mathbf{\nabla }^{2}\psi +%
\frac{b_{0}^{2}}{8m}\mathbf{\nabla }^{2}s_{\alpha }\cdot \left(
s_{\alpha
}-2\sigma _{\alpha }\right) \psi -\frac{b_{0}^{2}}{4m}\frac{\mathbf{\nabla }%
\rho }{\rho }\mathbf{\nabla }s_{\alpha }\sigma _{\alpha }\psi  \nonumber \\
&&-\left( 1-\frac{b_{0}^{2}}{\hbar ^{2}}\right) U_{\mathrm{B}}\psi
-V\left( x\right) \psi =0  \label{d6.14}
\end{eqnarray}%
where $U_{\mathrm{B}}$ is determined by the relation (\ref{d4.3}).
Deriving
dynamic equation (\ref{d6.14}), we have used the identities%
\[
\mathbf{s}^{2}\equiv 1,\qquad s_{\alpha }\mathbf{\nabla }s_{\alpha
}\equiv 0,\qquad \mathbf{\nabla }s_{\alpha }\left( \mathbf{\nabla
}s_{\alpha }\right) +s_{\alpha }\mathbf{\nabla }^{2}s_{\alpha
}\equiv 0
\]

\label{2beg}In the case if $b_{0}=\hbar $ the equation
(\ref{d6.14}) turns
into%
\begin{equation}
i\hbar \partial _{0}\psi +\frac{\hbar ^{2}}{2m}\mathbf{\nabla
}^{2}\psi -V\left( x\right) \psi +\frac{\hbar
^{2}}{8m}\mathbf{\nabla }^{2}s_{\alpha
}\cdot \left( s_{\alpha }-2\sigma _{\alpha }\right) \psi -\frac{\hbar ^{2}}{%
4m}\frac{\mathbf{\nabla }\rho }{\rho }\mathbf{\nabla }s_{\alpha
}\sigma _{\alpha }\psi =0  \label{d6.15}
\end{equation}%
where two last terms differ this equation from the Schr\"{o}dinger
equation. These two terms are responsible for vorticity of the
flow. In accordance with the Schr\"{o}dinger equation the particle
spin is an attribute of the quantum particle, and it does not
influence on the flow of the statistical ensemble. According to
the equation (\ref{d6.15}) a pointlike spin-free particle spin may
have a spin, generated by the vorticity of the statistical
ensemble flow. \label{2end}

Using the change of variables (\ref{d6.1}), (\ref{d6.3}), we did
not use the fact, that the solution of equations (\ref{d4.28}) is
a solution of the equations (\ref{d4.28b}). In the case of
description in terms of the wave function $\psi $ we have not the
problem, which we have at description in terms of the generalized
stream function $\mathbf{\xi }$\textbf{, }when
there are such solutions of (\ref{d4.28}), which are not solutions of (\ref%
{d4.28b}).

\section{Concluding remarks}

In this paper we try to construct the uniform formalism for
description of physical (stochastic and deterministic) systems. We
use the statistical ensemble as a basic object of dynamics, using
the fact that the statistical ensemble is a continuous dynamic
system independently of whether its elements are stochastic or
dynamic systems. Such an approach admits one to describe quantum
systems, considering them as stochastic system and using
\textit{only} principles of classical (not quantum) physics at
this description. Besides, the developed technique may be applied
for description of classical inviscid fluids.

\label{3beg}Existence of quantum particles together with the
uniform description of classical and quantum particles   generates
an alternative to quantum nature of particles in microcosm.
Indeed, the quantum paradigm supposes description of particle
motion by means of principles of quantum theory. The quantum
principles suppose a description in terms of linear dynamic
equations for a wave function, which is introduced axiomatically.
In the model conception, when the wave function is simply a way of
description of an ideal continuous media (statistical ensemble),
the dynamic equations appear to be linear only for nonrotational
flow of the medium. In the case of a rotational flow the
Schr\"{o}dinger equation ceases to be linear differential
equation. Nonlinear terms appear in the equation (\ref{d6.15}),
describing a statistical ensemble of quantum particles.

The uniform method of the particle motion description admits one
to refuse the quantum principles as needless ones. However, it
puts the question: "What is  the nature of the particle motion
stochasticity?" There is the only answer. The multivariance
(stochasticity) of the particle motion  is conditioned by the
properties of the space-time geometry. In the twentieth century,
when multivariant geometries were unknown, such an approach was
impossible and a use of the geometric paradigm was impossible. But
now, when multivariant geometries are known, the geometrical
paradigm, which denies the quantum principles as the prime
principles, looks more natural, than the quantum paradigm, based
on needless quantum principles, because a change of the space-time
geometry looks more reasonable, than a change of the dynamics
principles.\label{3end}.

We considered four different methods of the statistical ensemble
description. Consideration of deterministic, stochastic and
quantum systems as special cases of a physical system is
conditioned be the fact, that the space-time geometry may be
non-Riemannian, and the motion of particle may be multivariant
(stochastic) primordially \cite{R2010}.

Constructing uniform formalism, we did not introduce any new
hypotheses. We worked with physical principles (not with single
physical phenomena). We realized the logical reloading
\cite{R2010a}, i.e. replacement of basic concepts of a theory. A
single particle as a basic concept of the particle dynamics has
been replaced by another basic concept: statistical ensemble of
deterministic (or stochastic) particles. The logical reloading is
a logical operation, which is used rare in the theoretical
physics.

Usually the statistical description is used for a description of
particles, when information on the particle dynamics is
incomplete. This incompleteness may be connected with indefinite
initial conditions or with stochasticity of the particle motion.
Usually the statistical description is introduced as some external
operation with respect to the particle dynamics. The logical
reloading admits one to introduce the statistical description into
the particle dynamics. The statistical description becomes to be
an internal dynamical operation, which does not use the concept of
probability. Dynamical conception of statistical description, when
one considers many identical independent particles, but not a
probability of a state of a single particle, extends its
capacities, because the probabilistic description is only a
special case of the statistical description.

The logical reloading turns the statistical description into a
component of the particle dynamics. This circumstance extends
capacity of the particles dynamics. In particular, in the
relativistical case a description of the pair production (and
annihilation) becomes to be possible in the framework of the
uniform formalism of the particle dynamics.
\bibliographystyle{my-h-elsevier}

\end{document}